\documentclass[aps, prd, twocolumn, showpacs, a4paper,
nofootinbib, superscriptaddress,longbibliography, floatfix]{revtex4-1}
\usepackage{blindtext}
\usepackage{graphicx}
\usepackage{epsfig}
\usepackage{bm}
\usepackage{amsfonts}
\usepackage{pgf}
\usepackage{color}
\usepackage[T1]{fontenc}
\usepackage[latin1]{inputenc}
\usepackage[english]{babel}
\usepackage{amsmath}
\usepackage{amssymb}
\usepackage{amsfonts}
\usepackage{makecell}
\usepackage{hyperref}
\usepackage[draft,deletedmarkup=xout]{changes}
\usepackage{mathtools}
\usepackage{xcolor}
\usepackage{multirow}
\usepackage[version=4]{mhchem}
\usepackage{soul}
\usepackage{empheq}
\definecolor{green}{rgb}{0.19,0.64,0.54}
\definecolor{blue}{rgb}{0,0,1}
\definecolor{reddish}{rgb}{0.65, 0.2, 0.2}
\definecolor{darkgreen}{rgb}{0.2,0.7,0.3}
\definecolor{darkblue}{rgb}{0.3,0.40,0.48}
\definecolor{gray}{rgb}{.8,.8,.8}

\hypersetup{,
	pdftitle={CMB_CVOS},
  	pdfauthor={I. Rybak et al.},
    colorlinks=true,       
    linkcolor=darkblue,          
    citecolor=darkgreen,        
    filecolor=reddish,      
    urlcolor=blue,           
    linktocpage=true
}
\usepackage{wrapfig}
\usepackage{lipsum}
\usepackage{ulem}

\graphicspath{ {./Figures/} }

\newcommand{\dd}{\mathrm{d}}

\begin{document}

\title{Cosmic microwave background signatures from
    current-carrying cosmic strings}

\author{I. Yu. Rybak}
\email[]{irybak@unizar.es}
\affiliation{CAPA \& Departamento de F\'{\i}sica Te\'{o}rica, Universidad de Zaragoza, Pedro Cerbuna, 12, 50009 Zaragoza, Spain}
\affiliation{Centro de Astrof\'{\i}sica da Universidade do Porto, Rua
  das Estrelas, 4150-762 Porto, Portugal}
\affiliation{Instituto de Astrof\'{\i}sica e Ci\^encias do Espa\c co,
  CAUP, Rua das Estrelas, 4150-762 Porto, Portugal}

\author{C. J. A. P. Martins}
\email{Carlos.Martins@astro.up.pt}
\affiliation{Centro de Astrof\'{\i}sica da Universidade do Porto, Rua
  das Estrelas, 4150-762 Porto, Portugal}
\affiliation{Instituto de Astrof\'{\i}sica e Ci\^encias do Espa\c co,
  CAUP, Rua das Estrelas, 4150-762 Porto, Portugal}

\author{Patrick Peter}
\email{patrick.peter@iap.fr}
\affiliation{${\cal G}\mathbb{R}\varepsilon\mathbb{C}{\cal O}$ --
  Institut d'Astrophysique de Paris, CNRS \& Sorbonne Universit\'e,
  UMR 7095 98 bis boulevard Arago, 75014 Paris, France}
\affiliation{Centre for Theoretical Cosmology, Department of Applied
  Mathematics and Theoretical Physics, University of Cambridge,
  Wilberforce Road, Cambridge CB3 0WA, United Kingdom}

\author{E. P. S. Shellard}
\email{E.P.S.Shellard@damtp.cam.ac.uk}
\affiliation{Centre for Theoretical Cosmology, Department of Applied
  Mathematics and Theoretical Physics, University of Cambridge,
  Wilberforce Road, Cambridge CB3 0WA, United Kingdom}

\begin{abstract}
We continue our studies of the evolution and cosmological
consequences of current-carrying cosmic string networks, described by
a charge-velocity-dependent one scale (CVOS) model. We present a
detailed calculation of the effects of these networks on the cosmic
microwave background (CMB), in the context of this model, and
specifically discuss how such current-carrying strings may be
distinguished from their uncharged (Nambu-Goto) counterparts by
current or forthcoming CMB data. We find that, under the CVOS
hypothesis, the constraints on current-carrying strings should not
differ much from those of their structureless counterparts in that
the impact on the CMB can at most be reduced by a factor of $\sim
25\%$. Nevertheless, the presence of a current and charge affects
the distribution of power among scalar, vector and tensor modes,
and also its distribution between small and large scales. It should
therefore be possible for future high-sensitivity CMB temperature and polarization experiments
to distinguish between the two types of strings.
\end{abstract}

\date{\today}
\maketitle

\section{Introduction}

Cosmic strings provide a valuable probe for investigating possible
high-energy scenarios in the early Universe
\cite{JeannerotRocherSakellariadou, SarangiTye,
  Allys2,PhysRevD.106.123515}. To accurately constrain high-energy
physics using cosmic strings, it is crucial to understand the
evolution of their networks in quantitative detail. In particular,
this requires a comprehensive understanding of all possible additional
degrees of freedom and structures, such as long-lasting currents
\cite{Witten:1984eb,DAVIS1995197,LAZARIDES1985123,
  Peter:1993tm,GaraudVolkov,AbeHamadaYoshioka}, often dubbed
superconducting---a nomenclature we prefer not to use here as it often
refers to long-range electromagnetic effects---and Y-junctions
\cite{Copeland_2004,CopelandKibble}, each of which can introduce
deviations from the conventional cosmic string evolution paradigm
\cite{Oliveira:2012nj,Martins:2014,Vieira:2016,PhysRevD.83.063525,
  PhysRevD.99.063516}. Naturally, any such deviations can also impact
their observational predictions and the constraints obtained based on them \cite{PhysRevD.106.043521,PhysRevD.94.063529,Charnock:2016nzm,  Auclair_2021,Auclair_2023,RybakSousa2}.

To gain insights into the evolution of these one-dimensional
topological defects, it is more common to appeal to field theory
simulations
\cite{PhysRevD.105.063517,PhysRevD.104.063511,PhysRevD.107.043507}. Another
approach involves simulating cosmic strings as infinitely thin
entities governed by the Nambu-Goto action
\cite{Martins:2005es,Ringeval:2005kr,Blanco-PilladoOlumShlaer}. However,
these approaches have, so far, not successfully achieved full-scale
simulations of current-carrying string networks, although efforts
toward this goal are presently ongoing. Thus, to address the
evolution of a network of such strings, we make use of the
thermodynamical approach known as the charge-velocity-dependent
one-scale (CVOS) model \cite{Martins:2020jbq,Martins:2021cid,Rybak:2023jjn},
which extends the original VOS model developed for ordinary cosmic
strings in Refs.~\cite{Martins:1996jp,Martins:2000cs}.

A particularly sensitive and well-established probe to assess the
presence of cosmic strings is the study of cosmic microwave background
(CMB) anisotropies
\cite{PhysRevD.75.065015,Ringeval2010,PhysRevD.104.023507}, which rely on
linear physics, to which one may also add spectral distortions \cite{RUDAK1987346, CyrChlubaAcharya,CyrChlubaAcharya2} and lensing \cite{SazhinKhlopov, Sazhin2,
  Benabed:2000tr}.  In particular, anisotropies generated by ordinary
cosmic strings, whether they are estimated from field theory, Nambu-Goto
simulations or analytical approaches, yield similar predictions
\cite{LizarragaUrrestillaDaverioHindmarshKunz,LazanauShellard}; see
however Ref.~\cite{LazanuShellardLandriau} for detailed comparison and
possible discrepancies between different methods.

In the absence of simulations depicting the evolution of current-carrying string networks, albeit recent advancements in simulating individual superconducting strings \cite{Battye:2021sji,Battye:2021kbd, Fujikura:2023lil, Hiramatsu:2023epr}, we proceed to expand upon the \small{CMBACT} code \cite{PogosianVachaspati}. We employ the CVOS model to compute the CMB anisotropies resulting from these networks, while considering various assumptions regarding the model's phenomenological parameters. Previous work \cite{Rybak:2023jjn} has shown that, depending
on these model parameters and also the cosmological epoch, these networks
may find themselves in frozen regimes (where the charge and current dominate)
or in standard scaling solutions (with the charge and current disappearing).
Other things being equal, the former scenario may have some likelihood of happening in the radiation
era, while the latter is strongly indicated in the matter era. Once accurate simulations
of these networks become available one may expect to measure these parameters
directly from simulations, thereby determining which specific scenarios are
realized in practice, but for the moment we scan the parameter space, and
illustrate the results by considering representative examples of the various
possible scenarios and exploring the consequences of each of them.
Our results reveal distinctive CMB features associated with
the presence of charges and currents, which have the potential to
differentiate between general current-carrying and ordinary string
networks.

Throughout this work we use the following cosmological (Planck best
fit) parameters~\cite{Planck:2018vyg}: $\Omega_\text{b0} h^2=0.0224$,
$\Omega_\text{m0} h^2=0.1424$, $H_0 = 100 h \,
\text{km}\cdot\text{s}^{-1} \cdot \text{Mpc}^{-1}$, where
$h=0.674$. For definiteness, and in agreement with the currently
accepted constraints, when a fiducial value of the string tension is
necessary for a quantitative calculation, we set it to the value
$G \mu_0= 10^{-7}$ \cite{Planck2013}.

\begin{figure*}[t]
\begin{center}
\includegraphics[scale=0.38]{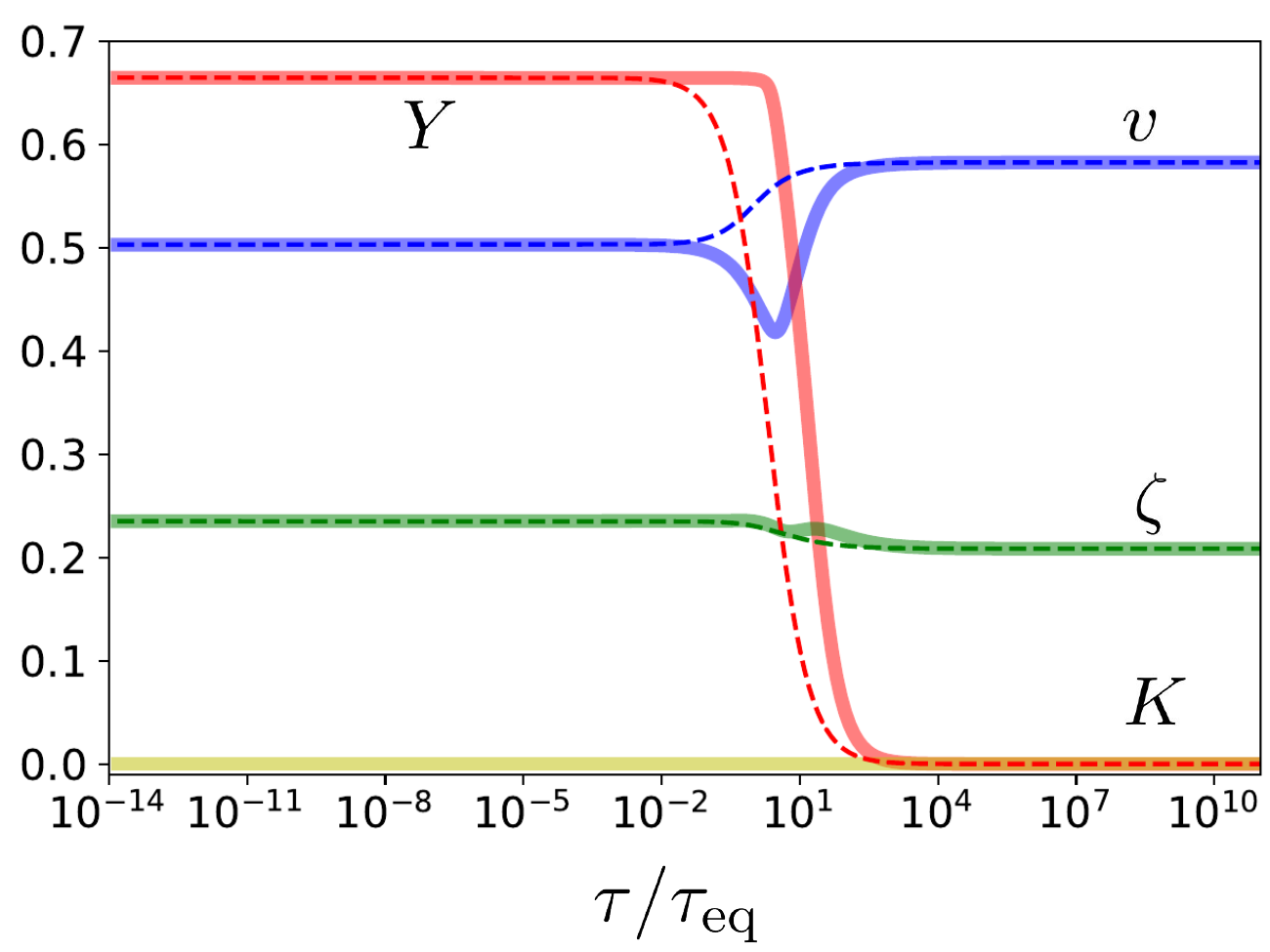}
\includegraphics[scale=0.38]{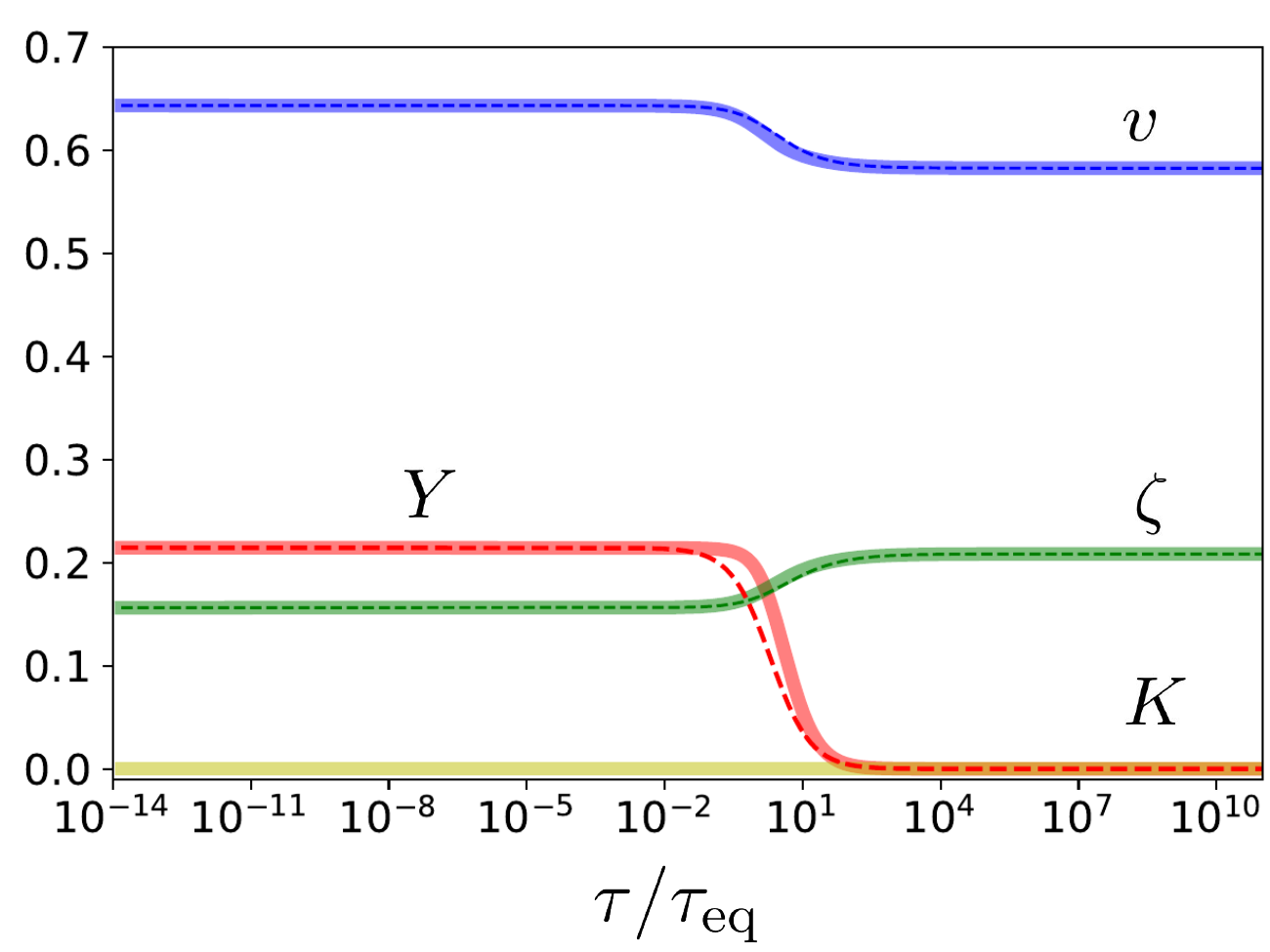}

\caption{ Evolution of the CVOS variables using the full system
    \eqref{EqOfMotMacroLin} (thick solid transparent lines) and the
    one-parameter approximation \eqref{EqOfMotMacroLinF} (dashed
    lines). The exact case is solved using the $\alpha-$dependent
    leakage function $A(Y)$ given by \eqref{Yfunction}, which allows
    the value for $Y_\textsc{sc}$ to be set in \eqref{Y_guess} as required
    for solving \eqref{EqOfMotMacroLinF}. Two different time evolutions
    are shown with different mass ratio parameters, namely $\alpha=1$
    (left panel) and $\alpha=3$ (right panel). Note that in both
    cases, differences are only visible around the radiation-matter
    transition, whose influence in the cosmological consequences is
    negligible, thereby justifying the one-parameter approximation.
\label{Figure:YS}
}

\end{center}
\end{figure*}

\section{CVOS formalism and linear model}

Current-carrying string networks can be described by the CVOS model
originally developed in Ref.~\cite{Martins:2020jbq}. We briefly review
its salient features in this section, referring the reader to the
original work for additional technicalities and discussion.

This model contains four key macroscopic parameters, which are:
\begin{itemize}
\item[(i)] the root-mean-square (RMS) velocity $v$ of the string network, that is the average magnitude of string velocities using a convenient statistical measure,
\item[(ii)] the characteristic length $\xi_\textsc{c}$ of the network, which represents the averaged distance between cosmic strings and can be related to the energy density of strings, excluding charge and current, as $$\rho_0 = \frac{\mu_0}{\xi_\textsc{c}^2},$$ where the constant $\mu_0$ is the tension or  mass per unit length of bare strings, 
\item[(iii)] the charge amplitude $$Y=\frac12 \left( Q^2+J^2\right),$$ with $Q$ and $J$ respectively the timelike and spacelike components of the averaged four-current, and
\item[(iv)] The chirality $$K=Q^2-J^2,$$ which determines whether the average string network possesses a globally timelike $K>0$, spacelike $K<0$ or lightlike $K=0$ (chiral) current. It can be expressed as an averaged Lorentz-invariant two-current amplitude. 
\end{itemize}
Another important ingredient
that is required in the CVOS model for a current-carrying cosmic
string network is an equation of state. The Witten model, considered
at first in Ref.~\cite{Witten:1984eb}, is characterized by the
following averaged equation of state \cite{Peter:1992dw}
\begin{equation}
\begin{cases}
\label{RealModel}
F_{\text{mag}}(K) = 1 - \displaystyle\frac{1}{2} \frac{K}{1 - \alpha
  K} & \text{for } K \leq 0, \\ & \\ F_{\text{elec}}(K) = 1 +
\displaystyle\frac{\ln (1- \alpha K) }{2 \alpha} & \text{for } \; K
\geq 0,
\end{cases}
\end{equation}
where the dimensionless parameter
$$\alpha = \left( \frac{m_\textsc{h}}{m_\sigma}\right)^2$$ is the
square ratio of the mass of the string-forming Higgs field
$m_\textsc{h}$ with the vacuum mass of the current-generating
condensate $m_\sigma$; this depends on the particular characteristics
of the symmetry breaking process which resulted in the formation of
current-carrying cosmic strings. In all of these
scenarios, we have $1<\alpha<\infty$ and,  while many cases of
particle physics relevance have $\alpha \gg 1$ holds, the strongest backreaction occurs with $\alpha \approx {\cal O}(1)$.

\begin{figure*}[t]
\begin{center}
\includegraphics[scale=0.9]{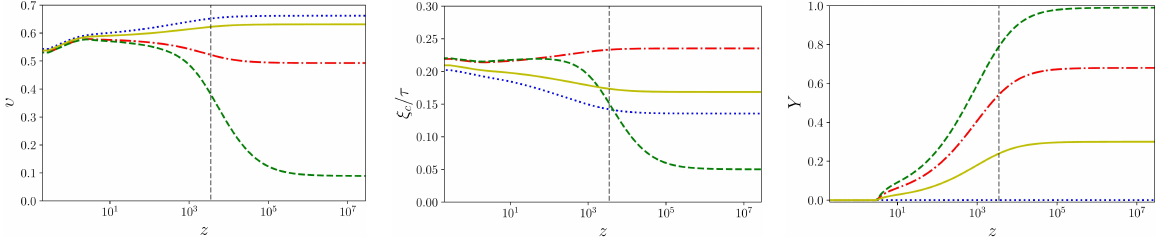}

\caption{\label{Fig:Y1Y2Y3z} Evolution of current-carrying cosmic
  strings thermodynamical quantities $v(z)$ (left panel),
  $\epsilon(z)$ (middle) and $Y(z)$ (right) as functions of the
  redshift $z$ for three different values of the scaling radiation
  charge, namely $Y_{\rm sc} =0.99$ (green dashed), $Y_{\rm sc} =0.68$
  (red dot-dashed) and $Y_{\rm sc} =0.3$ (yellow solid). These are
  compared with the uncharged Nambu-Goto evolution (blue dotted).
  Note that contrary to Fig.~\ref{Figure:YS} which was calculated
  using the radiation and matter solution \eqref{radmat}, this
  figure, as the \small{CMBACT} calculations below, is drawn using the
  full solution including the present cosmological constant
  dominated era. This explains in particular the sudden drop
  in the total charge for redshifts of a few.}

\end{center}
\end{figure*}

\begin{figure}
\begin{center}
\includegraphics[scale=0.55]{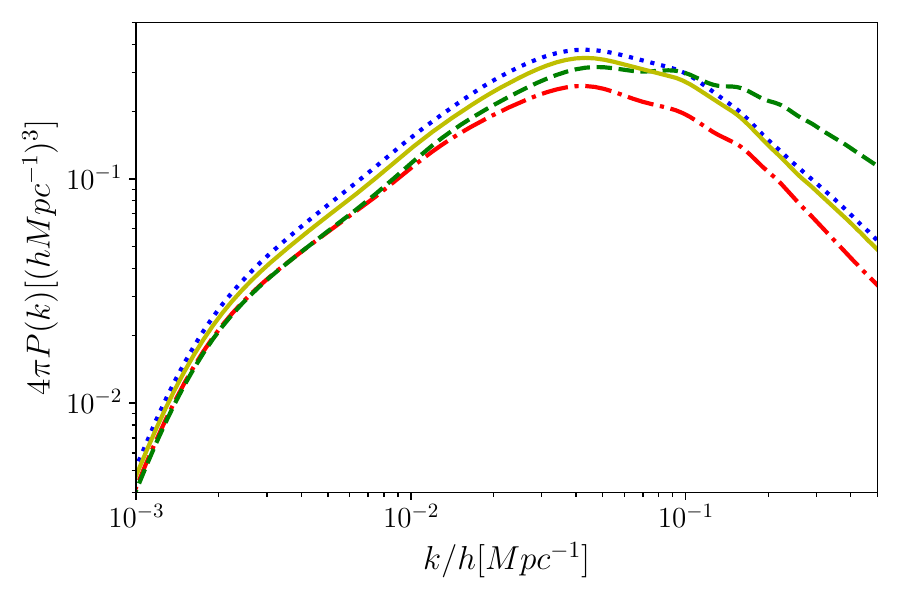}

\caption{\label{Fig:Y1Y2Y3Pk} Power spectra predicted for
  current-carrying cosmic strings with the same values and conventions
  as in Fig.~\ref{Fig:Y1Y2Y3z}.  The predicted power spectrum is not a
  simple function of the scaling radiation charge $Y_\textsc{sc}$, the
  intermediate value $Y_{\rm sc} =0.68$ (red dot-dashed line) leading
  to less power than the lowest one $Y_{\rm sc} =0.3$ (yellow
  solid). On small scales, the power is significantly larger for the
  highest value $Y_{\rm sc} =0.99$ (green dashed line). }

\end{center}
\end{figure}

We are interested in cosmic string networks evolving in an expanding
universe background defined by the homogeneous and isotropic
Friedman-Lema\^itre-Robertson-Walker (FLRW) metric
\begin{equation}
    \label{FLRW}
    \dd s^2 = a(\tau)^2 \left( \dd \tau^2 - \dd \bm{x}^2 \right),
\end{equation}
where $a(\tau)$ is the scale factor, $\tau$ is conformal time and
$\bm{x}$ represents spatial components.  The study of the CVOS model
with the equation of state \eqref{RealModel} on a FLRW background was
carried out in Ref.~\cite{Rybak:2023jjn}, The results therein suggest that for
most physically justified scenarios, in particular when
$K(0)\approx0$, and if one is interested in a macroscopic description,
the Witten model is identical or dynamically driven to the linear model
studied in Ref.~\cite{Martins:2021cid}.  Thus, in what follows we focus our
attention on the linear version of the CVOS model whose equations of motion read
\begin{equation}
\left\{ \begin{aligned}
\frac{\dd \xi_\textsc{c}}{\dd \tau} & = 
\frac{ \xi_\textsc{c} v^2 }{1 + Y}
\frac{\dot a}{a} 
+ \frac{ c v }{2 } + \frac{Y}{1+Y}v k(v), \\
 \frac{\dd v}{\dd \tau} & = \frac{(1-v^2)}{1 + Y}
\left[ \frac{k(v) \left(1 - Y \right) }{ \xi_\textsc{c} }
- 2 v \frac{\dot{a}}{a}  \right],  \\
 \frac{\dd Y}{\dd \tau} & =  2 Y \left[
\frac{v k(v)}{ \xi_\textsc{c} } - \frac{\dot{a}}{a}
\right] -  \frac{ Y}{\xi_\textsc{c} } A(Y),
 \end{aligned} \right.
\label{EqOfMotMacroLin}
\end{equation}
with the salient point that the chirality evolution equation
 \begin{equation}
\frac{\dd K}{\dd \tau}  = 2 K \left[
\frac{v k(v)}{ \xi_\textsc{c} } - \frac{\dot{a}}{a}
\right] - \frac{ K}{\xi_\textsc{c} } A(Y),
\label{EqOfMotMacroLinK}
\end{equation}
decouples from that of the other thermodynamical variables where we employed the comoving characteristic length $\xi = \xi_\textsc{c} a$ for convenience. When
numerically solving Eqs.~\eqref{EqOfMotMacroLin}, we set the chopping
efficiency to its Nambu-Goto (NG) value, i.e., $c\to
c_\textsc{ng}\simeq 0.23$ [except for our so-called Model (a), to
be defined in what follows, which has $c \to 0.5$], and the momentum parameter
$k(v)$ is assumed to be, following~\cite{Martins:2000cs}
\begin{equation}
    \label{MomentK}
    k(v) \to k_\textsc{ng}(v) = \frac{2 \sqrt{2}}{\pi} \frac{1-8
      v^6}{1+8 v^6} \left( 1-v^2 \right) \left( 1 + 2 \sqrt{2} v^3
    \right)\,.
\end{equation}
Note that here we are making the assumption that the fact that $k(v)$ only
depends explicitly on velocity, which is valid in the structureless case, still
applies on the presence of charge and current (or some combination thereof,
such as chirality). In other words, the assumption is that there is no explicit
dependence of these additional degrees of freedom. There is clearly an indirect
dependence, since the presence of charge or current will affect the velocity.
Testing this assumption will require numerical simulations, and is therefore
left for future work.

Although this reproduces the features measured in the string network
simulations, it is often sufficient, in order to describe the scaling
solution, to give $k(v)$ a constant value, i.e. $k(v) \to k_\text{sc}
= k(v_\textsc{sc})$ since in particular the function \eqref{MomentK}
presents a plateau in the relevant regime in which the solutions are
obtained. This is the choice we made in particular for our
  special cases, denoted (a), (b), (c) and (d) below, in
  agreement with Ref.~\cite{Martins:2021cid}.  It should be noted at
this point that the so-called {\sl chiral limit} $K\to 0$,
corresponding to the case for which the charge and the current are
equal, is an attractor for this system. As a result, we shall restrict
attention to this case only, and replace the full system by a single
parameter $Y_\textsc{sc}$ in the forthcoming CMB analysis.

The charge leakage efficiency $A$ is another phenomenological
parameter, not derivable from the microscopic equations of motion.  It
is introduced in order to take into account all the microscopic
effects, such as large local curvatures, through which the charge and
current initially flowing along the strings may escape the long string
network \cite{AbeHamadaSajiYoshioka}. Moreover, both timelike and
spacelike currents are seen to saturate somewhat for critical values,
so the leakage $A$ is expected to diverge when $Y$ reaches a critical
value, denoted $Y_\text{cr}$ \cite{BARR1987146}. Various functional
forms for $A(Y)$ can be used, as was suggested in Ref.~\cite{Rybak:2023jjn};
in what follows, we will assume
\begin{equation}
    \label{Yfunction}
    A(Y) = \frac{A_\text{c}}{ 1 - \text{e}^{-\left( Y-Y_\text{cr}
        \right)^2} },
\end{equation}
where $A_\text{c}$ and
\begin{equation}
\label{defineYcr}
Y_\text{cr} = \frac{2}{3 \alpha} =
\frac{2 m_\sigma^2}{3 m_\textsc{h}^2}
\end{equation}
are constant parameters which depend on the properties of the original
field theory model. We can note at this stage that with $\alpha \gg
1$, the critical current is expected to be rather small in physically
relevant cosmologies, thereby once again justifying the use of the
linear model. The system \eqref{EqOfMotMacroLin} can be solved
  numerically using the leakage function \eqref{Yfunction}, resulting
  examples begin shown in Fig.~\ref{Figure:YS}, superimposed with the
  analytic one-parameter expansion to which we now turn.

\section{One-parameter approximation for charge and current evolution}
\label{Ysc}

The evolution of the model described by Eqs. \eqref{EqOfMotMacroLin}
has been studied in detail in Refs.~\cite{Rybak:2023jjn,
  Martins:2021cid}. There it was shown that, for $a(\tau) \propto
\tau^n$, a nonvanishing charge leakage leads to a scaling behavior,
i.e.,
\begin{equation}
    \label{StandScal}
    L_\textsc{c} = \zeta_{\textsc{sc}} \tau, \quad \xi_\textsc{c} =
    \epsilon_\textsc{sc} \tau, \quad v = v_\textsc{sc}, \quad
    Y=Y_\textsc{sc},
\end{equation}
where the ``${\rm sc}$'' subscripts refer to the scaling values, which
are constants. The energy density of the string network is given by
\begin{equation}
\label{energy density}
\rho = \frac{\mu_0}{a^2 L_\textsc{c}^2}, \quad \text{with} \quad
\xi_\textsc{c}^2 = (1+Y) L_\textsc{c}^2
\end{equation}
(see Ref.~\cite{Martins:2020jbq} for details).

\begin{figure*}
\begin{center}
\includegraphics[scale=0.55]{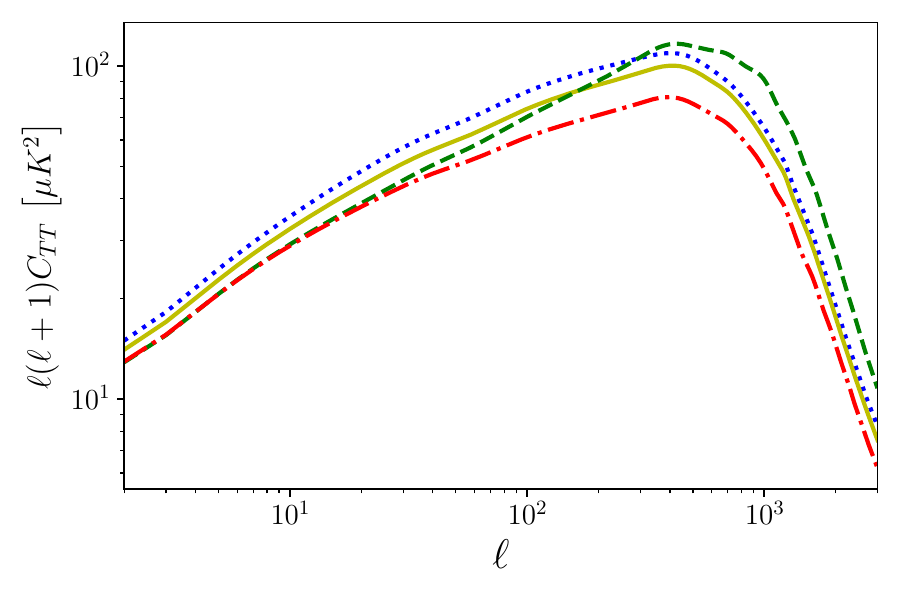}
\includegraphics[scale=0.55]{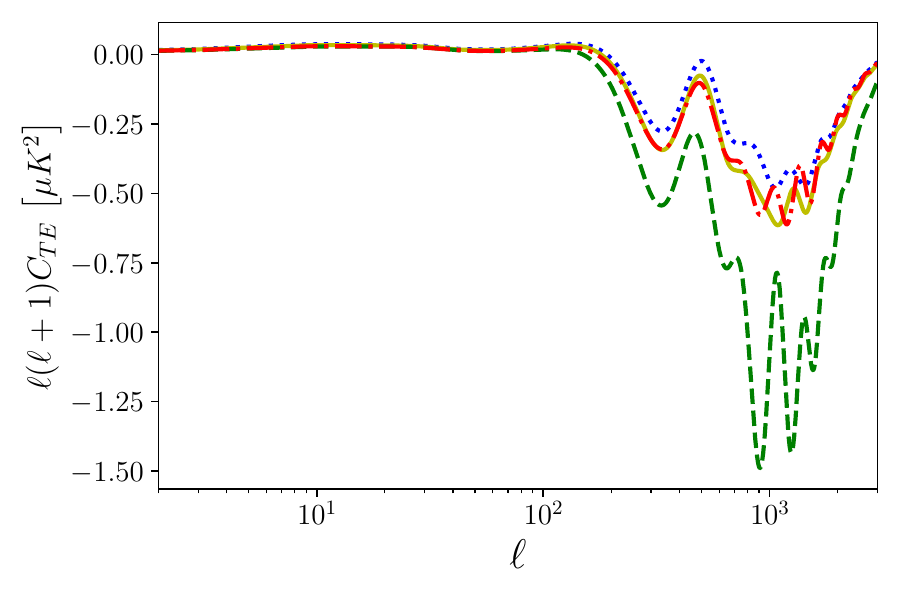}
\includegraphics[scale=0.55]{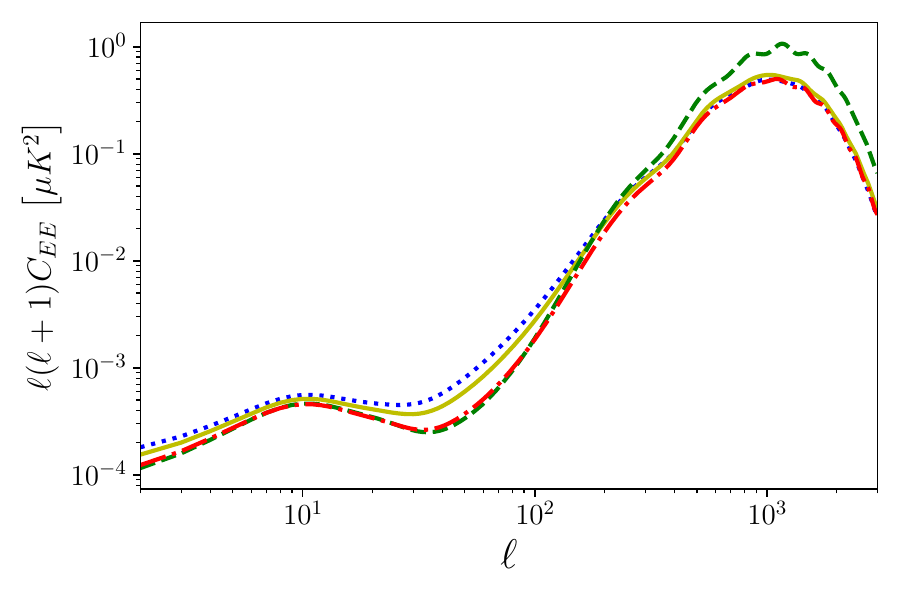}
\includegraphics[scale=0.55]{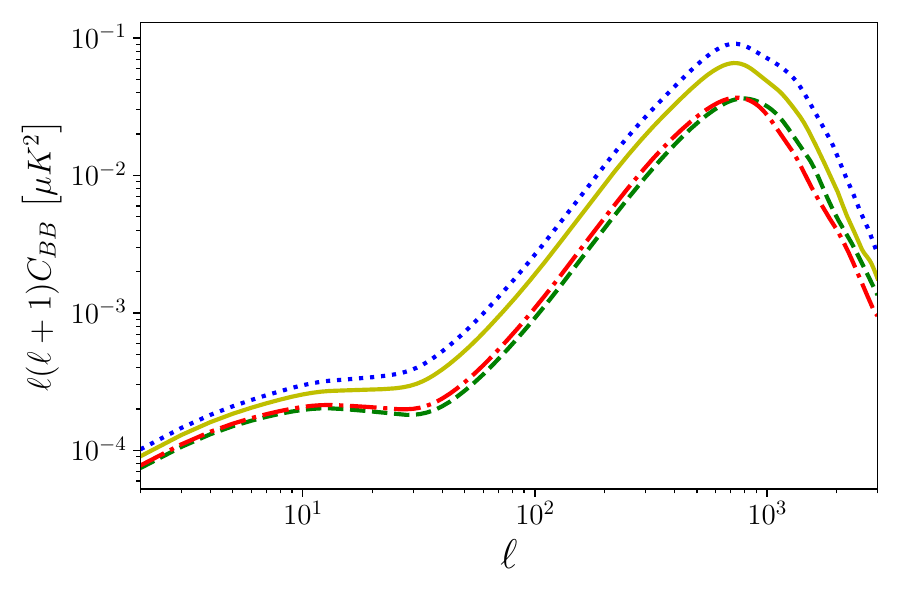}

\caption{\label{Fig:ClY1Y2Y3} CMB anisotropies predicted for a
  current-carrying cosmic string network in the different observable
  modes, namely $C_{_{TT}}$ (upper left panel), $C_{_{TE}}$ (upper
  right), $C_{_{EE}}$ (lower left) and $C_{_{BB}}$ (lower right), up
  to the normalization factors $\ell (\ell+1)$, as functions of
  the multipole $\ell$. These curves are shown for the same values and
  conventions as those in Fig.~\ref{Fig:Y1Y2Y3z} and represent the sum
  over all the contributions (scalar, vector and tensor)~--~details
  are presented in Appendix~\ref{AppA}. Although the largest
  effect comes from the largest scaling charge for the $TT$, $TE$ and
  $EE$ modes, the $BB$ mode contribution is larger for the Nambu-Goto
  case.  }

\end{center}
\end{figure*}

\begin{figure}
\begin{center}
\includegraphics[scale=0.55]{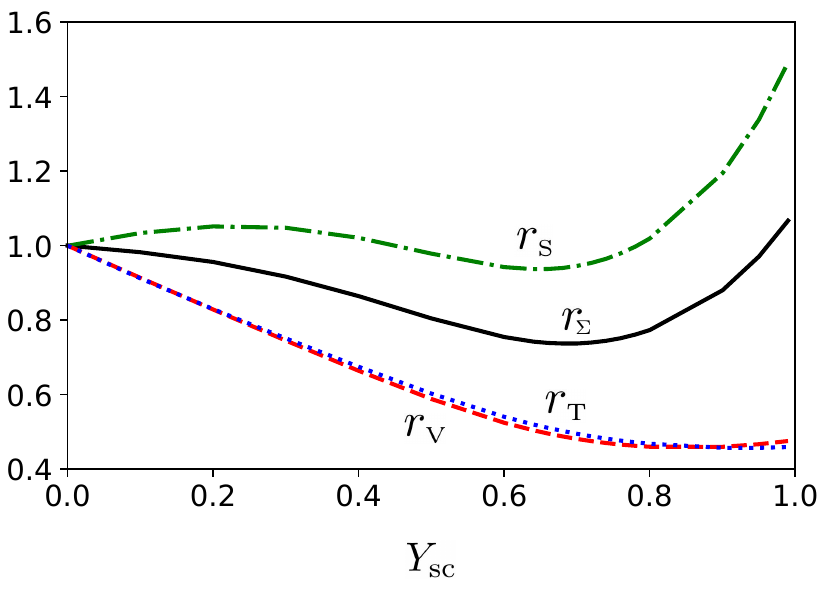}

\caption{\label{Fig:Ratios} The ratios $r_\textsc{i} = \max \left[
    C^\textsc{i}_{_{TT}}(Y_\textsc{sc})/C_{_{TT}}(0) \right]$, with
  $\textsc{i} = \textsc{s}$ (scalar contribution), $\textsc{i} =
  \textsc{v}$ (vectors), and $\textsc{i} = \textsc{t}$ (tensors)
  between the maxima of the relevant modes for the current-carrying
  case and the corresponding values for a Nambu-Goto network. The sum
  over all modes is shown as $\textsc{i} = \Sigma$. At most, one finds
  that the total amplitude can be reduced by a factor of around $25\%$
  so that inclusion of current-carrying capabilities in a cosmic
  string network cannot substantially modify the existing constraints
  on the string energy scale.}

\end{center}
\end{figure}

\begin{figure*}[t]
\begin{center}
\includegraphics[scale=1.3]{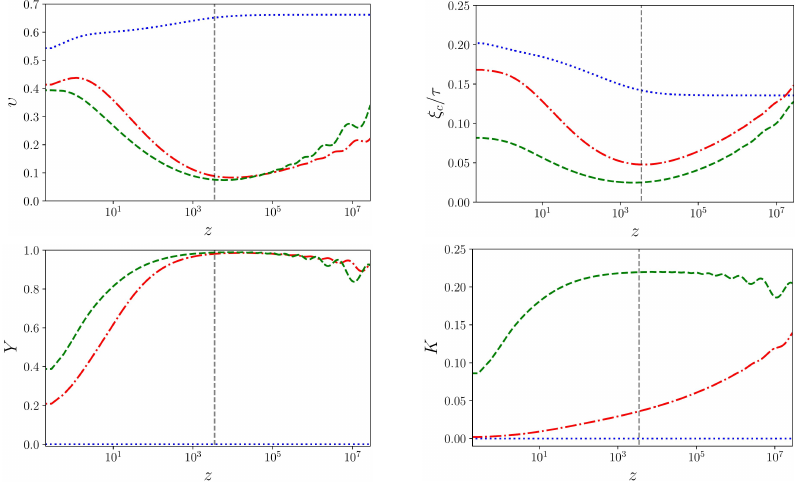}

\caption{\label{Fig:AC} Evolution of the thermodynamical quantities
  $v(z)$, $\epsilon(z)$ and $Y(z)$ as function of the redshift $z$ for
  the models denoted (a) (dot-dashed red line) and (c) (dashed
  green) in Ref.~\cite{Martins:2021cid} as well as the standard
  Nambu-Goto evolution (dotted blue) for comparison. Both models have
  vanishing leakage $A_\text{c}=0$, constant momentum parameter
  $k_\text{(a)} = 0.6$ and $k_\text{(c)}=0.4$, and bias
  $g_\text{(a)}= 0.9$ and $g_\text{(c)}=1$, model (a)
  having a slightly higher than Nambu-Goto chopping efficiency
  $c_\text{(a)}=0.5$. The absence of leakage lead to so-called frozen
  configurations in which the RMS velocity $v$ and correlation length
  decrease in the radiation era. In both cases, the charge $Y$
  saturates during the radiation era. The chirality takes a nonvanishing scaling value in the radiation era for model (c) but it falls monotonically for model (a). Both models also have a nonvanishing charge throughout the matter era, albeit decreasing.}

\end{center}
\end{figure*}

\begin{figure*}[t]
\begin{center}
\includegraphics[scale=1.3]{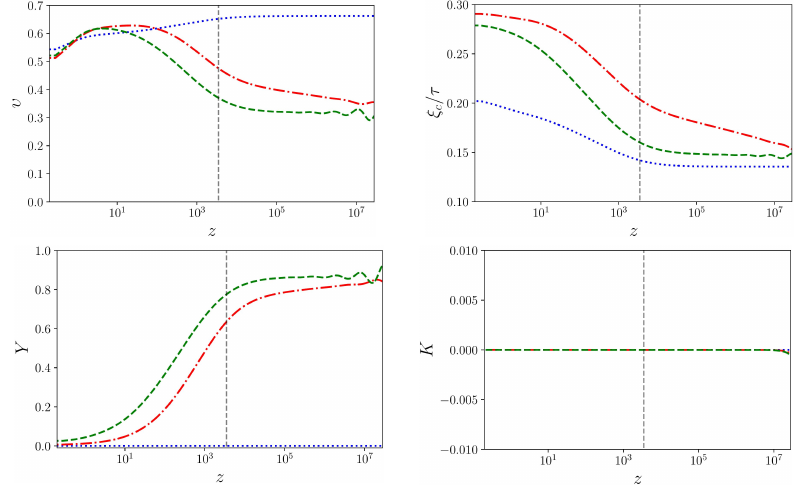}

\caption{\label{Fig:BD} Evolution of the thermodynamical quantities
  $v(z)$, $\epsilon(z)$ and $Y(z)$ as functions of the redshift $z$
  for the models denoted (b) (dot-dashed red line) and (d) (dashed
  green) in Ref.~\cite{Martins:2021cid} as well as the standard
  Nambu-Goto evolution (dotted blue) for comparison. Model (b) has
  no leakage ($A_\text{c}=0$) and a bias function (see text)
  $g = 1+1.2 Y$,
  while model (d) has no bias and a large constant leakage
  $A(Y)=0.6$. Both
  models assume a constant momentum parameter $k = 0.7$ and can be
  seen to belong to the radiation scaling cases.}

\end{center}
\end{figure*}

Throughout the radiation dominated epoch, depending on initial
conditions, the CVOS model predicts two possible evolution regimes,
namely the standard Nambu-Goto evolution with trivial (vanishing)
charge $Y_\textsc{sc} = 0$, and a current-carrying (or charged) regime
with $Y_\textsc{sc} \neq 0$. This is to be contrasted with the matter
dominated epoch (and any faster expansion rates), which has only one
possible regime, which is a
standard Nambu-Goto evolution with $Y_\textsc{sc} = 0$. Thus, the
evolution of the system \eqref{EqOfMotMacroLin} in an expanding
universe is determined only by a charge amplitude $Y_\textsc{sc}$ in
the radiation epoch.
 
The actual value of $Y_\textsc{sc}$ depends on the type of cosmic
  string under consideration (in particular, it should be a function
  of the parameter $\alpha$) and the available physical mechanisms for
  the current leakage. It follows that we can simplify the system
  \eqref{EqOfMotMacroLin} by assuming that the charge $Y$ evolves
  phenomenologically as
\begin{equation}
    Y_{\rm f} = Y_\textsc{sc} \left( 2 - \frac{\dot{a}}{a} \tau
    \right) 
    =\frac{2Y_\textsc{sc}\tau_\mathrm{eq}}{\tau + 2\tau_\mathrm{eq}},
    \label{Y_guess}
\end{equation}
where the exact solution \cite{Mukhanov:1990me}
\begin{equation}
    a(\tau) = a_\mathrm{eq} \left[ 2
      \left(\frac{\tau}{\tau_\mathrm{eq}} \right) +
      \left(\frac{\tau}{\tau_\mathrm{eq}} \right)^2 \right],
\label{radmat}
\end{equation}
connecting the radiation and matter dominated eras, was used.

By changing $Y_\textsc{sc}$, we obtain generic predictions that
  allow us to distinguish current-carrying from ordinary cosmic
  strings. Thus, in what follows we treat $Y_\textsc{sc}$ as a free
  parameter that spans all possible scenarios of CMB anisotropies
  generated by a chiral current-carrying cosmic string network. In
  that case the CVOS together with \eqref{Y_guess} can be written as
\begin{equation}
\left\{ \begin{aligned}
 \frac{\dd \xi_\textsc{c} }{\dd \tau} & =
 \frac{ \xi_\textsc{c} v^2 }{1 + Y_{\rm f}}
 \frac{\dot a}{a} + \frac{ c v }{2 } + 
 \frac{Y_{\rm f}}{1+Y_{\rm f}}v k(v), \\
 \frac{\dd v}{\dd \tau}  & =  \frac{(1-v^2)}{1 + Y_{\rm f}}
\left[ \frac{k(v) \left(1 - Y_{\rm f} \right) }{ \xi_\textsc{c} }
- 2 v \frac{\dot{a}}{a}  \right].
\end{aligned} \right.
\label{EqOfMotMacroLinF}
\end{equation}
Solving the above system using the value of $Y_\textsc{sc}$ provided
by the solution of the full system \eqref{EqOfMotMacroLin} allows us to
assess the validity of the approximation. Such a comparison is shown
as an illustration in Fig.~\ref{Figure:YS}.

To compare the one-parameter model with the complete evolution depicted in Fig.~\ref{Figure:YS}, we can use Eq.~\eqref{Y_guess}. However, when dealing with \small{\small{CMBACT}}, it becomes necessary to incorporate a Heaviside function, denoted as $\Theta(\dots)$, to prevent the charge value from reaching negative values: 
\begin{equation}
    Y_{\rm f} = Y_\textsc{sc} \left( 2 - \frac{\dot{a}}{a} \tau
    \right) \Theta \left( 2 - \frac{\dot{a}}{a} \tau \right).
    \label{Y_guess2}
\end{equation}

\begin{figure}[t]
\begin{center}
\includegraphics[scale=0.55]{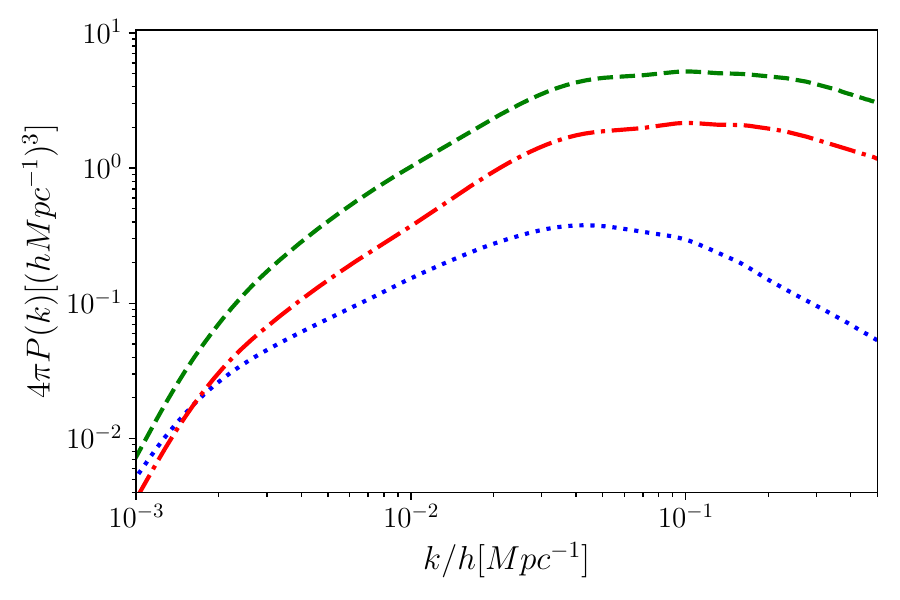}

\caption{\label{Fig:ACPk} Power spectra for current-carrying
  cosmic strings described by the so-called frozen
  models (a) and (c). The parameter
  choices and color conventions follow the same notation as in
  Fig.~\ref{Fig:AC}.}

\end{center}
\end{figure}

\begin{figure}[t]
\begin{center}
\includegraphics[scale=0.55]{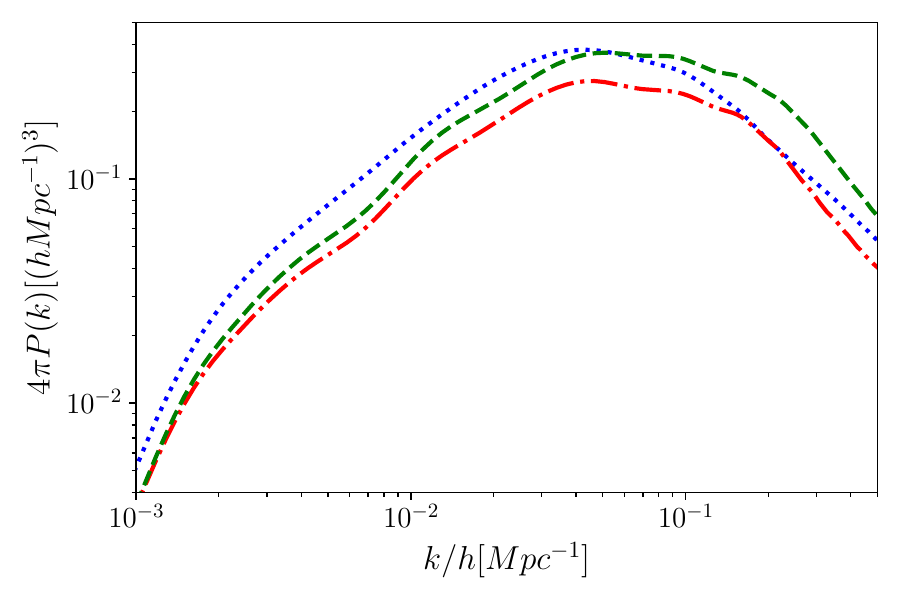}

\caption{\label{Fig:BDPk} Power spectra for current-carrying cosmic
  strings described by (b) and (d) models. The parameter choices and color conventions follow the same notation as in
  Fig.~\ref{Fig:BD}.}

\end{center}
\end{figure}

Despite the small deviation of the $Y$ function close to the
radiation-matter transition period, the simple model described by
Eqs.~\eqref{EqOfMotMacroLinF} reproduces the CVOS time development
quite well throughout most of the evolution of the universe.
Using this simplified model, we are able to scan the relevant
parameter space by varying the value of a single parameter,
$Y_\textsc{sc}$ (cf. Sec.~\ref{single} below). Using a modified
  version of \small{CMBACT}~\cite{PogosianVachaspati} applied to the CVOS
  model, as briefly described in Appendix ~\ref{AppA}, we can obtain the
  cosmological consequences expected from a current-carrying cosmic
  string network in the CMB.


\subsection{The single-parameter model.} \label{single}

As previously explained, we focus our analysis on the model with
a single parameter $Y_\textsc{sc}$ described by the simplified system
\eqref{EqOfMotMacroLinF} with \eqref{Y_guess2}. It is then sufficient to vary the parameter
$Y_\textsc{sc}$ and calculate how it impacts the CMB anisotropy
predictions. Substituting the dynamics of the system under consideration described by Eqs.~\eqref{EqOfMotMacroLinF} as functions of the redshift rather than time, shown in
Fig.~\ref{Fig:Y1Y2Y3z}, the stress-energy
tensor components required for the \small{CMBACT} code
\cite{PogosianVachaspati} are found from the expressions 
\eqref{Stress-Energy} for the different
values of $Y_\textsc{sc}$. The scalar-vector-tensor
decomposition contributions from the string network are shown in Fig.~\ref{Fig:CMB_Y} in Appendix ~\ref{AppA}, while the resulting late-time matter power spectrum $P(k)$ is shown in Fig.~\ref{Fig:Y1Y2Y3Pk}.  The angular power spectrum of CMB anisotropies are shown in Fig.~\ref{Fig:ClY1Y2Y3}, which depicts the temperature  ($T$) and $E-$ and $B-$polarization
correlations $C_{_{TT}}$, $C_{_{TE}}$, $C_{_{EE}}$ and $C_{_{BB}}$.  

The CMB and dark matter perturbation spectra result from the interplay of scalar, vector and tensor contributions from strings moving at relativistic velocities, which are more complex than inflationary fluctuations dominated by adiabatic scalar modes. The subtle effect of altering the magnitude of  string currents is illustrated in the matter power spectrum $P(k)$ shown in Fig.~\ref{Fig:Y1Y2Y3Pk}.
A small scaling charge makes
hardly any difference compared with the NG case, as expected, but when the charge increases and there is significant reduction in the string velocity, then there is less power on all scales, although
predominantly at small ones.  At this point, the primary perturbative effect of strings remains through their motion, creating ``wakes'' in the dark matter distribution or discontinuities in the CMB temperature \cite{Stebbins:1987cy, NevesdaCunha:2018urr}. 

Increasing the scaling charge to $Y_\textsc{sc}=0.99$ causes the velocity to drop even further with the network density rising higher due to the smaller spacing between cosmic strings at these larger charges. In this case, the charged string power spectrum in Fig.~\ref{Fig:Y1Y2Y3Pk} begins to exceed the 
NG prediction on small scales.  This effect is also reinforced by the reduced string velocity because there is a cross-over between the gravitational effects of the string motion and the scalar Newtonian potential generated by the charge along the string (absent for a straight NG string).   In general, the maximal amplitude of the power spectrum decreases for current-carrying cosmic strings and moves toward smaller length scales.  This phenomenon could conceivably hold intriguing implications for early structure formation, as proposed in Ref.~\cite{PhysRevD.108.043510}, establishing some of the necessary conditions for the formation of massive black holes at high redshifts originating from current-carrying strings, as discussed in Ref.~\cite{10.1093/mnras/stac1939}.

The various CMB correlation functions are shown in
Fig.~\ref{Fig:ClY1Y2Y3}, and they demonstrate a similar nontrivial effect
with the increase of the scaling charge $Y_\textsc{sc}$. The temperature autocorrelation starts decreasing with the charge, with an overall
amplitude drop a factor which we found to be at most
of the order $25\%$; this could presumably lead to a slight easing of the cosmic strings 
constraint from current CMB observations. On the other hand, for larger charges there is a cross-over which leads to larger temperature anisotropies on smaller scales, as noted already for the matter power spectrum. The relative contributions of the scalar, vector and tensor modes to the temperature autocorrelations are shown in Fig~\ref{Fig:Ratios}.

The $TE$ cross correlation is hardly affected at all by the presence
of a current flowing along the long strings, expect for very high
charges. The same conclusion holds for the $EE$ polarization, though an enhancement of roughly a factor of two can be observed at high charge.
Finally, the distinctive string network $BB$ correlation appears to be a roughly monotonically
decreasing function of the scaling charge, halting at a plateau above
which its maximum amplitude is half of the NG prediction.  The $BB$ contribution from vector and tensor modes is due to the string velocity which is suppressed by the presence of large charges.

\begin{figure*}
\begin{center}
\includegraphics[scale=0.55]{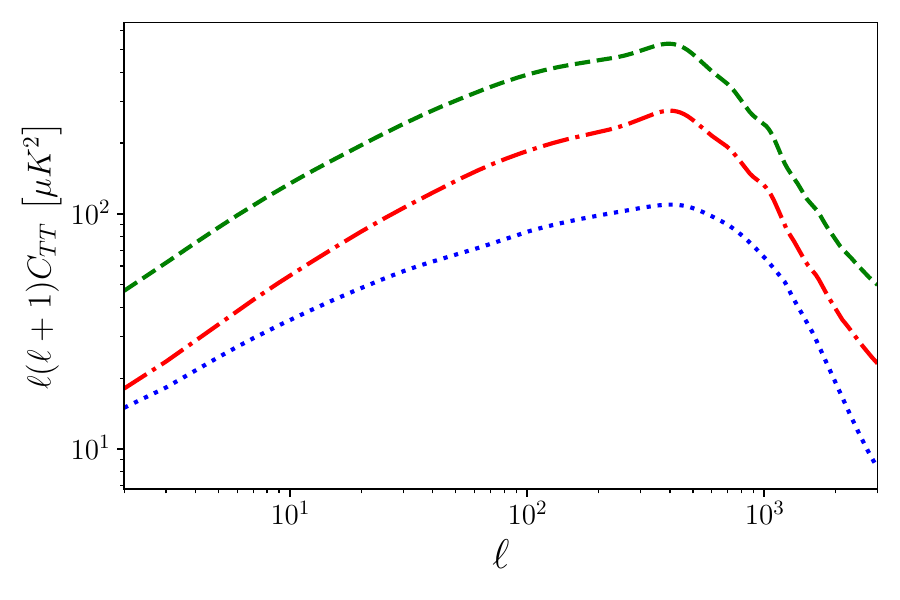}
\includegraphics[scale=0.55]{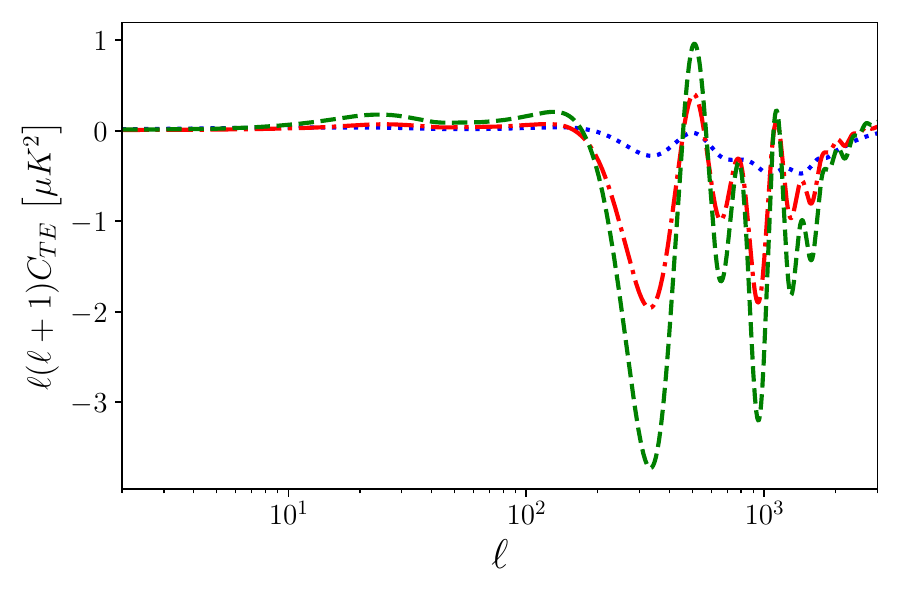}
\includegraphics[scale=0.55]{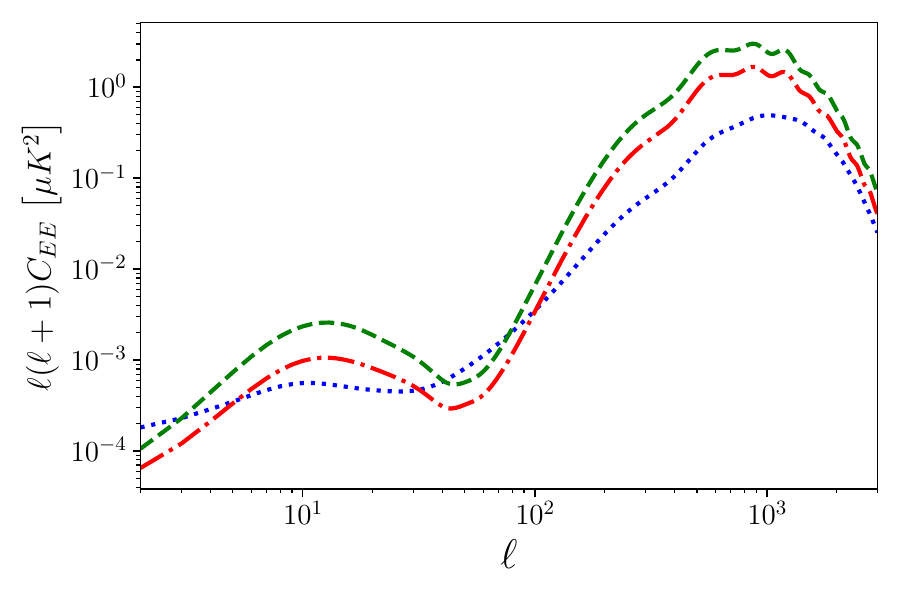}
\includegraphics[scale=0.55]{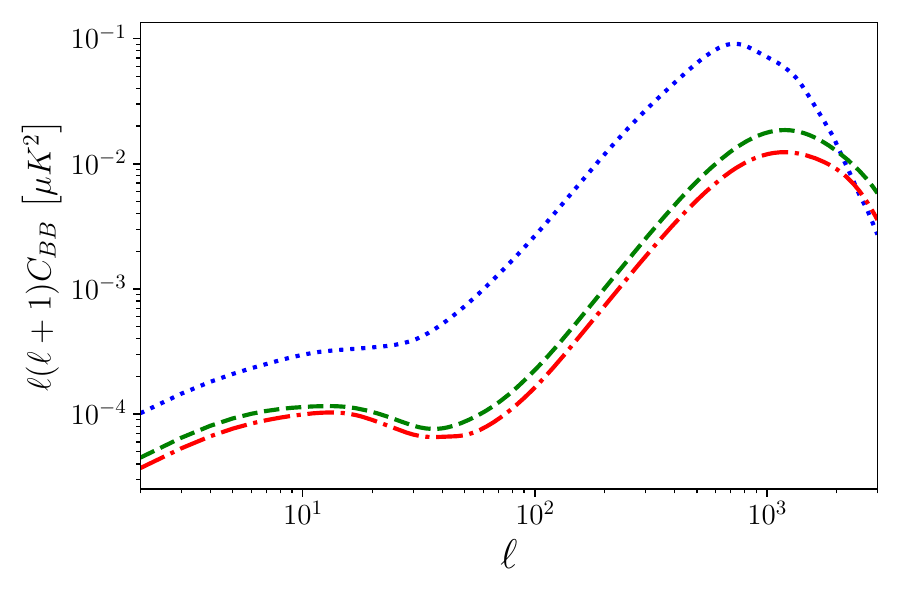}

\caption{\label{Fig:ClAC} CMB anisotropies predicted for a from
  current-carrying cosmic string network in the different observable
  modes, namely $C_{_{TT}}$ (upper left panel), $C_{_{TE}}$ (upper
  right), $C_{_{EE}}$ (lower left) and $C_{_{BB}}$ (lower right), up
  to the normalization factors $\ell (\ell+1)/(2\pi)$, as functions of
  the multipole $\ell$. These curves are shown for the same value and
  color convention as those in Fig.~\ref{Fig:AC} and represent the sum over
  all the contributions (scalar, vector and tensor). Further details are
  presented in Appendix ~\ref{AppA}. Dot-dashed red lines represent model (a), dashed
  green lines - model (c) as defined in Ref.~\cite{Martins:2021cid} and dotted blue lines - Nambu-Goto.}

\end{center}
\end{figure*}

\begin{figure}
\begin{center}
\includegraphics[scale=0.43]{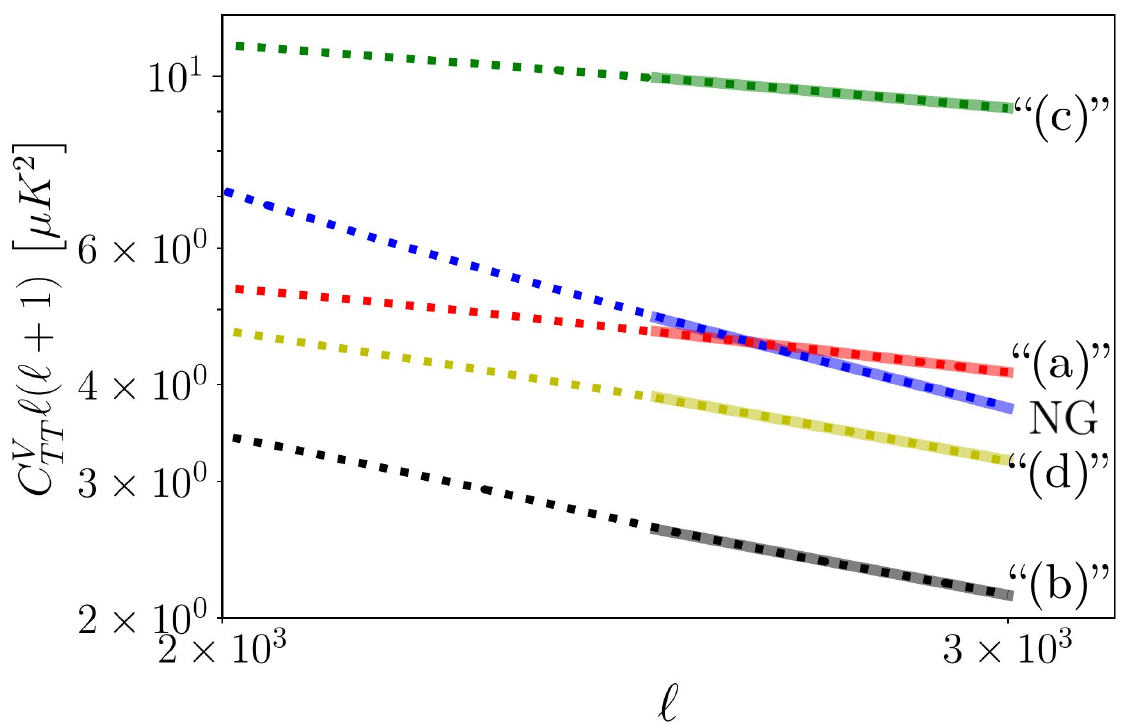}
\caption{\label{Fig:VectFit} Fit of vector modes (represented by straight lines) as the dominant $\ell>800$ component for CMB anisotropies $C^V_{TT}(\ell)$ (represented by dots) produced by current-carrying cosmic string networks described by models (a) to (d) models , as well as the Nambu-Goto case (NG).  
 The best-fit exponents \eqref{ACfit} for models (a) to (d) are as follows: $-0.67$, $-1.1$, $-0.5$, and $-1$, with $-1.5$ for NG.} 

\end{center}
\end{figure}

\subsection{Other relevant scenarios}
\label{Other scenarios}

On top of the above models, described mainly by a single parameter,
Ref.~\cite{Martins:2021cid} (and Fig.~1 therein) identified 4 particular
models, unimaginatively called (a), (b), (c) and (d), whose redshift
evolution is displayed in Figs.~\ref{Fig:AC} and \ref{Fig:BD}.  These
models represent different scenarios of current-carrying string
network evolution, and the aforementioned work discusses how they span
the possible range of physically realistic evolution scenarios. Let us
now give a brief description for each case.

The models described above assume no bias in the loop chopping efficiency,
with respect to charge and current, but this can also be included. Specifically,
when a NG long
string network loses energy through loop formation, the
chopping efficiency $c$ is defined by~\cite{Martins:1996jp}
\begin{equation}
\left. \frac{\dd E_\textsc{ng}}{\dd \tau} \right|_\text{loops} = 
-c v \frac{E_\textsc{ng}}{\xi_\textsc{c}},
\label{chopNG}
\end{equation}
while for a current-carrying network, this relation
becomes~\cite{Martins:2021cid}
\begin{equation}
\left. \frac{\dd E}{\dd \tau} \right|_\text{loops} = 
- g(Y,K) c v \frac{E}{\xi_\textsc{c}},
\label{chop}
\end{equation}
thereby introducing the bias function $g$ which {\it a priori}
depends on both the charge $Y$ and the chirality $K$. The
single-parameter model of the previous section shares
with the NG scenario the unbiased property, in that it
assumes $g=1$, contrary to the particular cases which we now discuss.

More details about these four models can be found in Ref.~\cite{Martins:2021cid}.
The redshift developments of the frozen category of
models (a) and (c), with behavior \eqref{Frozen},
are shown in Fig.~\ref{Fig:AC},
while the more regular (b) and (d) scenarios 
with linear scaling behavior \eqref{StandScal} are
shown in Fig.~\ref{Fig:BD}. For convenience, we show the
evolution of the rms velocity $v$, charge amplitude $Y$
and conformal characteristic length $\xi^2_\textsc{c} =
\mu_0 (1+Y)/(\rho a^2)$ (with $\rho$ the string network density)
for frozen networks in Fig.~\ref{Fig:AC}. In particular,
we see that during the radiation period there is a power
law decay of $v$ and $\xi_\textsc{c}/\tau$ according to 
Eq.~\eqref{Frozen}, while the charge amplitude $Y$ tends
to its saturated value $Y_\text{sat} = 1$ as a power law
as well. One can see that the choice of parameters in the
model (c) leads to much smaller values of $\xi_\textsc{c}
/\tau$ in comparison with the model (a).

\subsubsection{Model (a): Frozen chiral}

The first particular case of interest
we studied in Ref.~\cite{Martins:2021cid} corresponds to what we
called a frozen string network. Assuming no (or negligible) losses,
i.e. setting $A_\text{c}=0$ in Eq.~\eqref{Yfunction}, the charge
tends to saturate to the constant value $Y_\textsc{sc} \to 1$
while the velocity decreases to vanishingly small values, hence
the ``frozen'' network qualification. In this model, the
momentum parameter is given a constant value, in practice we
chose $k(v) = k_\text{(a)} = 0.6$, which is compatible with the fact
that as the velocity decreases, $k(v)$ varies very little
and eventually reaches a plateau for $v\ll 1$. We
also set the chopping efficiency $c$ to the
value $c_\text{(a)} = 0.5 > c_\textsc{ng}$, slightly larger than the
value obtained in NG simulations, and assumed a constant bias
$g(Y,K) = g_\text{(a)} = 0.9$: the value of the bias parameter $g$
controls whether loops typically lose above ($g>1$) or below
($g<1$) the average current and charge from the network.

These assumptions lead to nontrivial current during the whole string
network evolution, and indeed, we find that even during the
matter dominated era, the charge gets a nonvanishing value. 
In the absence of any relevant
charge leakage mechanism in this
scenario, the string network is frozen during the radiation period,
and we found that the thermodynamical quantities evolve according 
to~\cite{Martins:2021cid}
\begin{equation}
\label{Frozen}
v \propto \frac{\xi_\textsc{c}}{\tau} \propto 
\sqrt{1-Y} \propto \tau^{-s},
\end{equation}
where $s$ is a positive constant: during the radiation-dominated era, the
long string network stops moving and it becomes denser, with the correlation
length decreasing.
The matter dominated epoch is characterized by the approach to scaling of the network, as defined in Eq.~\eqref{StandScal}, including a nontrivial charge $Y\neq0$ value.

This scenario is also denoted ``chiral'' because although the
charge $Y$ saturates to unity, this number being a consequence
of our choice of the linear equation of state (it can be above
or below in the nonlinear regime~\cite{Rybak:2023jjn}), the
chirality $K$ rapidly decreases even if given an initial 
nonvanishing value.

\subsubsection{Model (b): Linear scaling}

This type of model shows a nontrivial scaling charge $Y\neq0$ in the
radiation epoch, i.e. the full scaling behavior of Eq.\eqref{StandScal}, 
followed by an ordinary Nambu-Goto string network with
vanishing charge $Y=0$ in the matter dominated epoch. In
this case, looking rather similar to the one-parameter
model of Sec.~\ref{single}, the bias parameter $g$ of
Eq.~\eqref{chop} is made into a function of the charge
(we restrict attention to a linear behavior as a first guess), namely
\begin{equation}
g=1+b Y,
\label{bY}
\end{equation}
and we set a positive value for the constant $b$. As in
the previous case, we consider a negligible leakage
$A_\text{c}=0$ but a larger momentum parameter $k_\text{(b)}=0.7$.

\subsubsection{Model (c): Frozen nonchiral}

Changing only slightly
the parameters of model (a) yields a different situation
in which although the charge saturates to $Y\to 1$ in the
radiation dominated era, there is a nonvanishing chirality
(positive in this case). Here, the chopping efficiency
is set to its NG value $c_\textsc{ng}=0.23$, while the
momentum parameter is still assumed constant, but smaller,
namely $k_\text{(c)} = 0.4$, and there is no bias, i.e.
$g_\text{(c)}=1$. In this case, the nonchiral and charged
radiation dominated era is followed by the approach to matter-dominated scaling
in which both the charge and the chirality decay to
negligible values (i.e. asymptotically the NG network scaling solution without charge or current).

\subsubsection{Model (d): Linear scaling with leakage}

The final special case is also rather close to the previous single-parameter case,
also exhibiting the standard scaling behavior of Eq. \eqref{StandScal}
with nontrivial charge $Y\neq0$ during the radiation dominated
epoch which vanishes after the radiation to matter domination
transition. The parameters are chosen such that the momentum
parameter is set $k_\text{(d)}= 0.7$ (larger than in (a) and (c)),  the chopping efficiency as for the NG case, 
and with no bias, but the key difference is that we assume
a constant leakage function $A(Y) = A_\text{(d)} = 0.6$.

\begin{figure*}
\begin{center}
\includegraphics[scale=0.55]{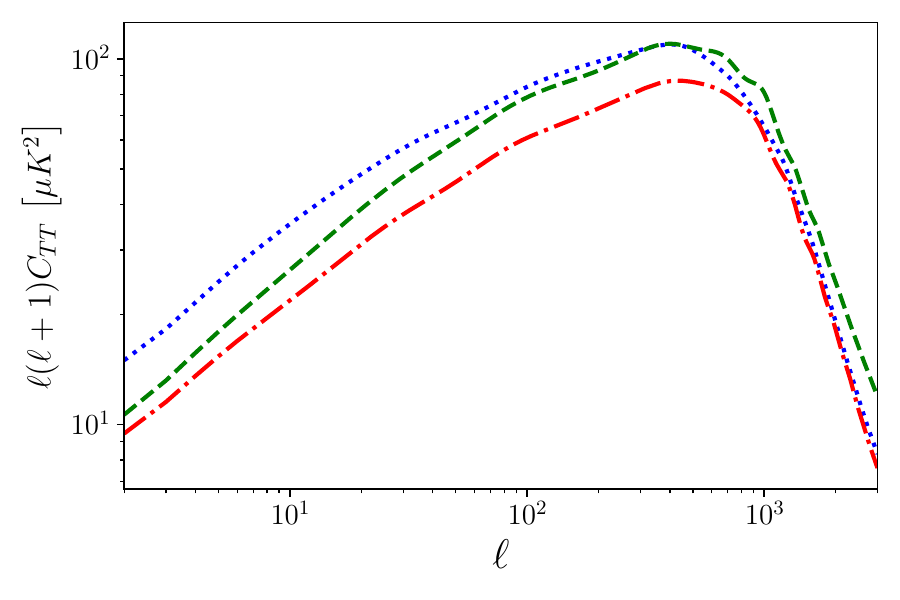}
\includegraphics[scale=0.55]{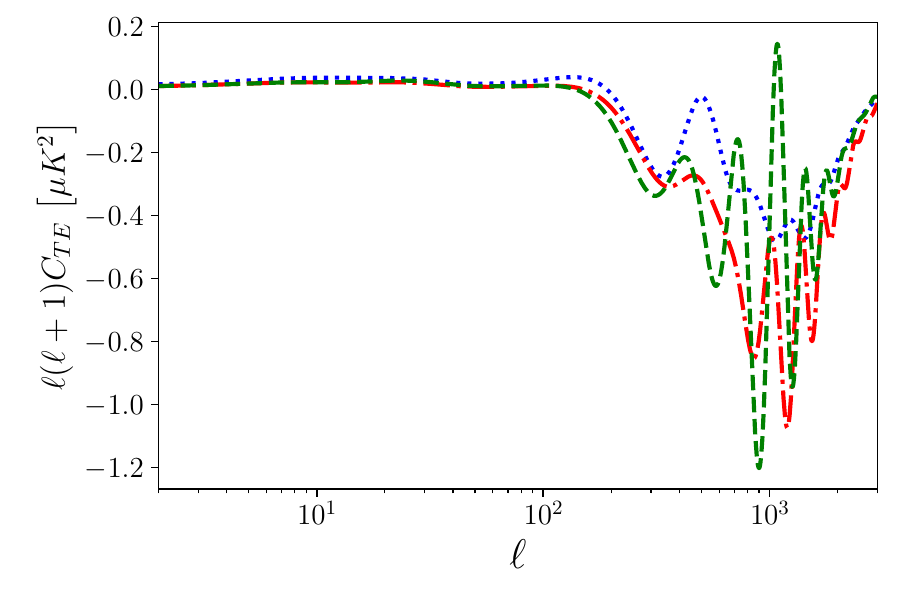}
\includegraphics[scale=0.55]{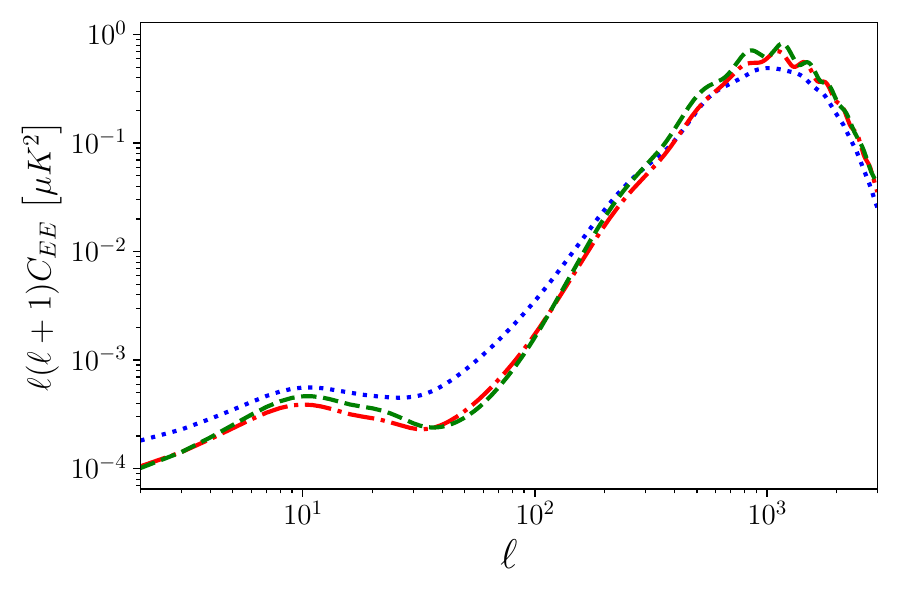}
\includegraphics[scale=0.55]{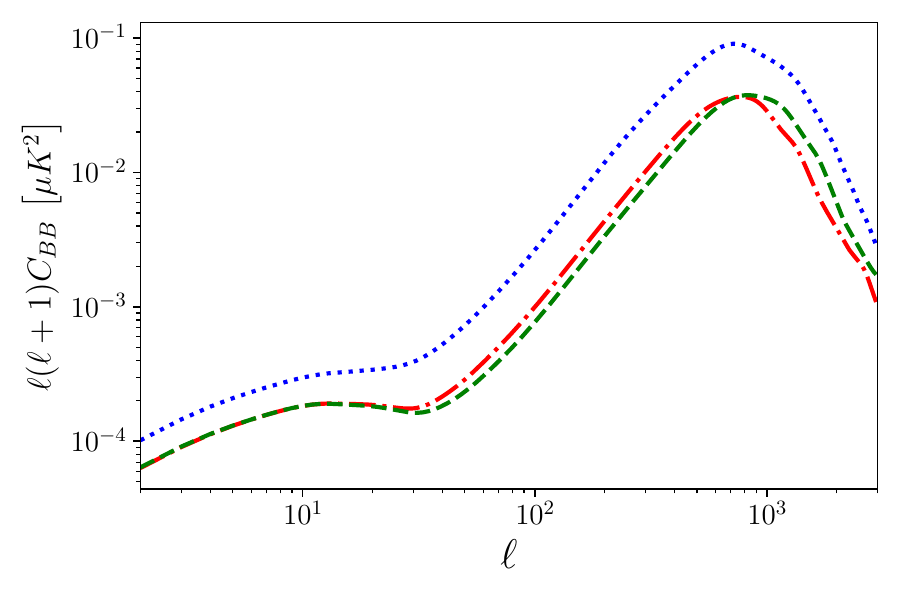}

\caption{\label{Fig:ClBD} CMB anisotropies predicted for a from
  current-carrying cosmic string network in the different observable
  modes, namely $C_{_{TT}}$ (upper left panel), $C_{_{TE}}$ (upper
  right), $C_{_{EE}}$ (lower left) and $C_{_{BB}}$ (lower right), up
  to the normalization factors $\ell (\ell+1)$, as functions of
  the multipole $\ell$. These curves are shown for the same value and
  convention as those in Fig.~\ref{Fig:BD} and represent the sum over
  all the contributions (scalar, vector and tensor)~--~details are
  presented in Appendix ~\ref{AppA}. Dot-dashed red lines represent model (b), dashed
  green lines - model (d) as defined in Ref.~\cite{Martins:2021cid} and dotted blue lines - Nambu-Goto.}

\end{center}
\end{figure*}

\begin{figure}[b]
\begin{center}
\includegraphics[scale=0.56]{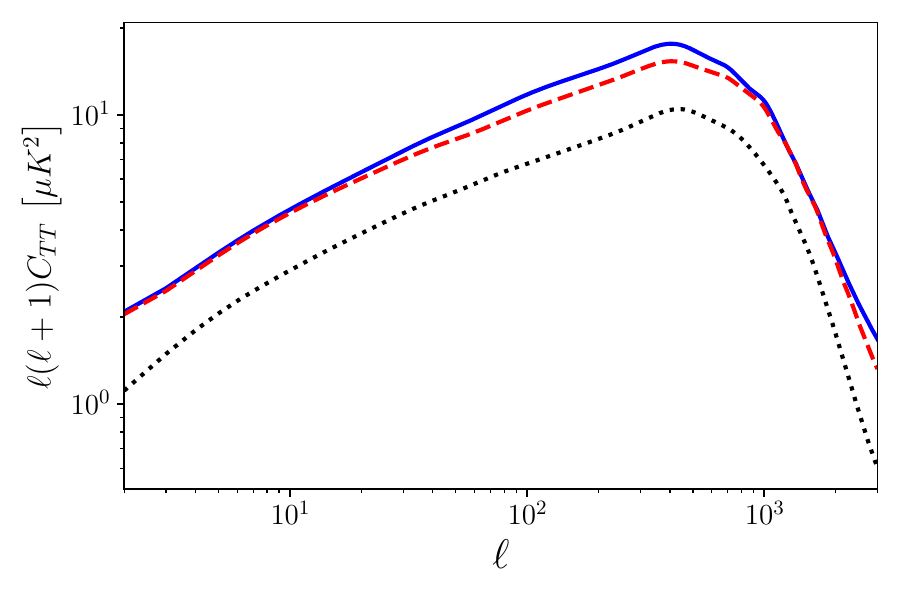}
\caption{\label{Fig:COMPARISON} Comparison between the $C_{TT}$ modes of the full model (solid blue line), described by Eqs.~\eqref{EqOfMotMacroLin}, and the single-parameter approximation (dashed red line), given by Eqs.~\eqref{Y_guess2} and \eqref{EqOfMotMacroLinF}. In the single-parameter model, we set $Y_{\rm sc} = 0.9$, whereas in the full model, we adjusted parameters to ensure that during the radiation period, the scaling value of the charge amplitude $Y_{\rm rd} = 0.9$. Additionally, for comparison purposes, we present the standard Nambu-Goto result with a black dotted line.} 

\end{center}
\end{figure}

\section{CMB predictions}

By using the Eqs.~\eqref{Stress-Energy} in the Appendix together with
the evolution of CVOS models which for the four CVOS model examples of
Ref.~\cite{Martins:2021cid} in the \small{CMBACT} code, we can now obtain the
corresponding CMB anisotropy predictions. 
The first conclusions that can be drawn concern the matter power
spectrum shown in Fig.~\ref{Fig:ACPk} for the cases
(a) and (c) and in Fig.~\ref{Fig:BDPk} for
 (b) and (d). Not unexpectedly, the
power spectrum from the models (b) and (d), whose
behavior is quite similar to that of the single-parameter
approximation, also predicts relatively small deviations
from the NG spectrum, with potentially less
power on large scales and more on small scales,  as shown in Fig.~\ref{Fig:BDPk}.
The frozen cases (a) and (c), on the other hand, show much more noticeable deviations
from the standard behavior, with an amplified spectrum across all scales, though most notably at small length scales, where  the discrepancy can reach the level
of one or two orders of magnitude. If such models are physically realistic, then 
very stringent constraints on the underlying parameters will ensue.

The resulting CMB anistropies for models (a) and (c) are illustrated in  Fig.~\ref{Fig:ClAC}.   The  larger scalar modes from model (c) yield larger $TT$ and $EE$ autocorrelations.  The minimum velocities for (a) and (c) examples are close
to each other, as are the $1-Y$ values which results in a similar magnitude for the vector and tensor modes, which are significantly reduced compared
to the Nambu-Goto case, as seen in the Appendix  Fig.~\ref{Fig:CMB_AC}.   On the other hand, the more dense model (c) leads
to a larger vector mode contribution than the model (a). Still, both contributions,
from the (a) and (c) models, reduce vector modes significantly in
comparison to the standard Nambu-Goto cosmic strings. The increase of
the contribution of current-carrying cosmic strings can be observed
only for high multipole moments $\ell>10^3$ of the vector modes.

The particular increase of vector modes for high multipole moments
$\ell$ occurs only for string networks that go through a frozen type
of evolution. The frozen network clearly changes the relation
$\ell(1+\ell) C_{_{TT}} \propto \ell^{-1.5}$ \cite{Fraisse:2007nu, LevonPogosian_2009}
for long strings. Carrying out the fit for $\ell>2500$, as illustrated in Fig.~\ref{Fig:VectFit}, we obtain that
the models (a) and (c) behave as
\begin{equation}
    \label{ACfit}
\ell(1+\ell) C^\text{(a)}_{_{TT}} \propto \ell^{-0.67}, \quad
\ell(1+\ell) C^\text{(c)}_{_{TT}} \propto \ell^{-0.50},
\end{equation}
providing a potential signal for small-scale anisotropy
\cite{LevonPogosian_2009}. However, one should keep in mind, that the presence of loops will
eventually dominate for $\ell \gtrsim 2000$, assuming the relation
$\ell(1+\ell) C^\text{loops}_{_{TT}} \propto \text{const.}$ obtained in
Ref.~\cite{PhysRevD.104.023507}.

 The CMB anisotropies for models exhibiting linear scaling regimes (b) and (d) are depicted in Fig.~\ref{Fig:ClBD}. As anticipated, these models demonstrate behavior akin to the single-parameter approximation. This similarity is visually evident when comparing Fig.~\ref{Fig:ClY1Y2Y3} with Fig.~\ref{Fig:ClBD}, indicating that the single-parameter approximation effectively captures the main features of the (b) and (d) models. This is also evidenced by direct comparisons between the CMB anisotropies obtained from this simplification and those from the full parameter model, as shown in Fig.~\ref{Fig:COMPARISON}.  The small differences observed, which do not exceed 15\% even for large current values $Y\sim 0.9$, suggest that the single-parameter approximation offers useful insights for the CMB anisotropies generated by current-carrying cosmic string networks, at least for those models exhibiting a linear scaling regime of evolution.

\section{Conclusions}

We have provided a first study of the CMB anisotropy for
current-carrying strings, relying on our previously developed CVOS
model.  We have shown that as the charge is increased, there is a
minimum contribution to the $TT$ correlation. In particular, one can see
the sum of $TT$ components in Fig.~\ref{Fig:ClY1Y2Y3}. The dashed line
in this plot, with $Y_\textsc{sc}=0.68$, is approximately the smallest
contribution that can be obtained for the $TT$ correlation due to the
presence of the current, according to the CVOS model with equation of
state \eqref{RealModel}. This minimal value can be also seen in
Fig.~\ref{Fig:Ratios}, where $r_\Sigma = \max \left[
  C_{TT}(Y_\textsc{sc})/C_{TT}(0) \right]$ has a minimum at
$Y_\textsc{sc} \approx 0.68$ with an amplitude reduction of around
25\%.

In Fig.~\ref{Fig:Ratios} we also see that the maxima for the vectors $r_\textsc{v} =
\max \left[ C^\textsc{v}_{TT}(Y_\textsc{sc})/C_{TT}(0)
  \right]$ and tensors $r_\textsc{t} = \max \left[
  C^\textsc{T}_{TT}(Y_\textsc{sc})/C_{TT}(0) \right]$
decreases monotonically as the current contribution increases. Hence,
the increase of the ratio of scalar to vector/tensor modes plays a characteristic role in the CMB anisotropy signal that can
distinguish standard Nambu-Goto strings from cosmic strings with
nontrivial internal structure.

As a final caveat, we note that our results are based on two
  levels of approximation. The first stems from the simplifying
  assumptions of piecewise source terms in the \small{CMBACT} Einstein-Boltzmann code. The second pertains to the CVOS
 network evolution model itself. Although this builds upon the standard VOS model,
  which has been the subject of extensive testing and calibration
  (including both field theory and Nambu-Goto simulations), the
  analogous process of calibration has not yet been achieved for the CVOS
  extension per se. The reason for this is simply the absence of
  reliable simulations of the evolution of string networks with
  charges and currents.  It is our understanding that such simulations
  are now becoming available, so a more detailed and robust study of
  the cosmological consequences of these networks will become possible
  in the near future.

It is essential to note that our analysis does not include certain potentially intriguing phenomena that could have an influence on the evolution of a current-carrying string network. Specifically, we have not incorporated the effect of primordial magnetic fields that may arise from an early universe phase transition as, for example, discussed in Ref.~\cite{VACHASPATI1991258}. In principle, such a magnetic field could generate currents along the strings, nontrivially modifying the  evolution of cosmic string networks (see, e.g., Ref.~\cite{PhysRevD.74.025012}). This fascinating avenue remains open for  further exploration.

\acknowledgments

This work was financed by Portuguese funds through FCT -
Funda\c{}c\~ao para a Ci\^encia e a Tecnologia in the framework of the
project 2022.04048.PTDC (Phi in the Sky, DOI
10.54499/2022.04048.PTDC) and R\&D project 2022.03495.PTDC (uncovering the nature of
cosmic strings). C.J.M. also acknowledges FCT and POCH/FSE (EC) support through Investigador FCT Contract No. 2021.01214.CEECIND/CP1658/CT0001 (DOI 10.54499/2021.01214.CEECIND/CP1658/CT0001). P.S. acknowledges funding from the
STFC Consolidated Grants No. ST/P000673/1 and No. ST/T00049X/1. I.R. also acknowledges support from the Grant No. PGC2022-126078NB-C21 funded by MCIN/AEI/ 10.13039/501100011033 and ``ERDF A way of making Europe'', as well as Grant No. DGA-FSE grant No. 2020-E21-17R from the Aragon Government and the European Union - NextGenerationEU Recovery and Resilience Program on `Astrof\'{\i}sica y F\'{\i}sica de Altas Energ\'{\i}as' CEFCA-CAPA-ITAINNOVA.

\appendix

\section{\small{CMBACT} for CVOS}
\label{AppA}

\begin{figure*}[t]
\begin{center}
\includegraphics[scale=0.78]{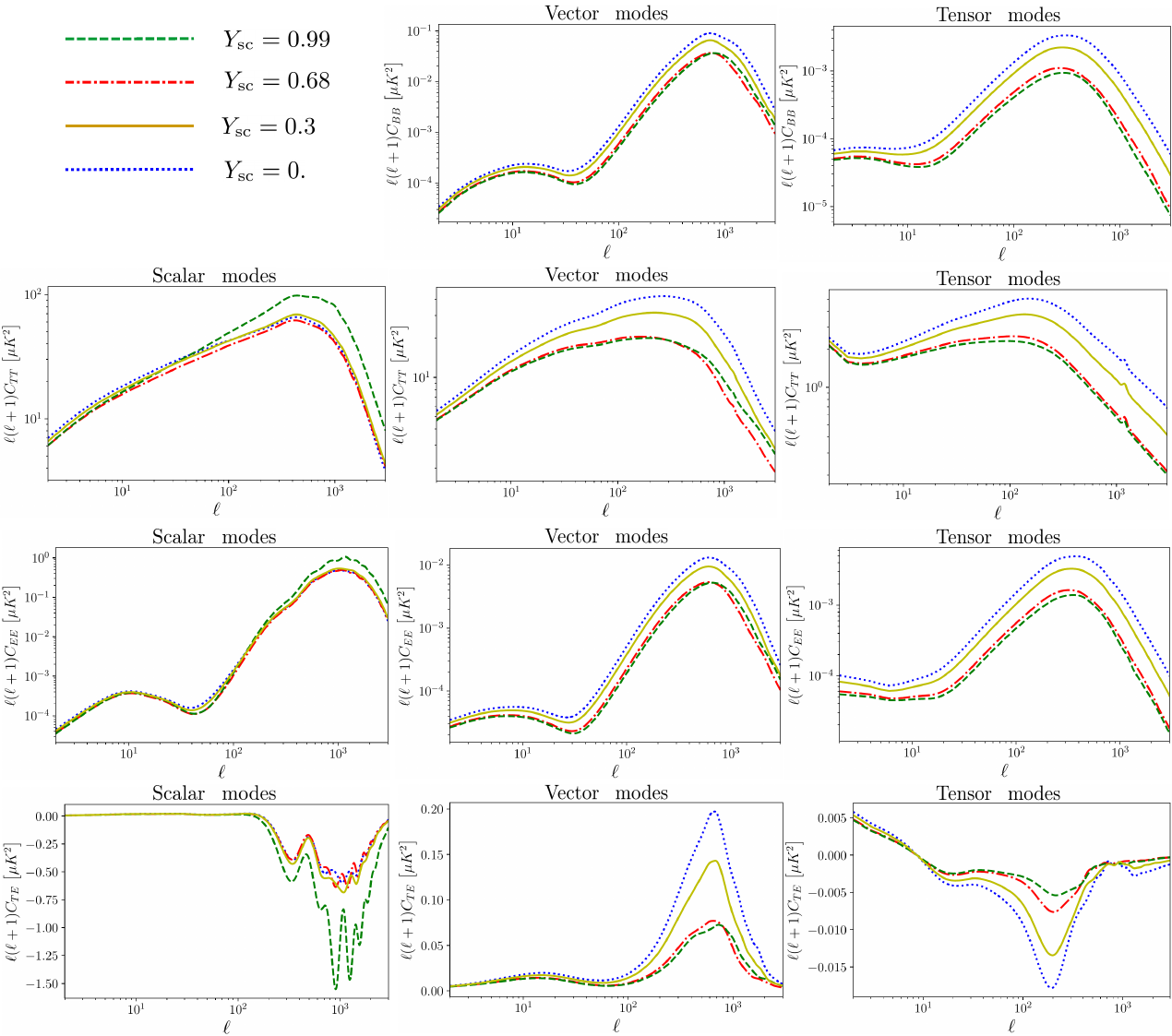}

\caption{\label{Fig:CMB_Y} The $BB$, $TT$, $EE$, and $TE$ components of the CMB anisotropies induced by current-carrying cosmic strings evolving according to Eqs.~\eqref{EqOfMotMacroLinF} are examined with varying values of $Y_\textsc{sc}$.}

\end{center}
\end{figure*}

We use a modified version of the publicly available \small{CMBACT} code
\cite{PogosianVachaspati} to describe the CMB anisotropies generated
by current-carrying cosmic strings, in comparison with standard
NG scenario. Our primary goal is to identify differences that
may lead to future discriminating tests between the two types of
strings.

Here we provide a concise description of the methodology employed
  in the \small{CMBACT} code \cite{PhysRevD.59.023508, PogosianVachaspati} for
  calculating anisotropies generated by cosmic strings, including the
  extension for current-carrying strings. The \small{CMBACT} code is built
  upon the CMBFAST linear Einstein-Boltzmann equations solver, which
  requires the stress-energy tensor of the active source as an
  input. Consequently, it is essential to obtain a Fourier-transformed
  stress-energy tensor that accurately represents the string
  network.

The \small{CMBACT} code utilizes randomly oriented straight segments to
  approximate a realistic string network. These straight segments have
  a length $\xi$ and move in random directions with a velocity $v$,
  gradually decaying while ensuring that their total energy
  corresponds to that of the VOS (or CVOS) model. Therefore, we can
  represent the collection of straight segments as follows
  ~\cite{PogosianVachaspati}:
\begin{equation}
\label{StrSeg}
X^{\mu} = x_0^{\mu} + \sigma \hat{X}^{\mu}_1 + \tau \hat{X}^{\mu}_{2},
\end{equation}
where
\begin{equation}
\label{UnitVectors1}
      \hat{X}^{\mu}_1 = \begin{pmatrix} 0 \\ \sin \theta \cos \phi
        \\ \sin \theta \sin \phi \\ \cos \theta \end{pmatrix}
\end{equation}
and
\begin{equation}
\label{UnitVectors2}
\hat{X}^{\mu}_2 = \begin{bmatrix} 1 \\ \upsilon ( \cos \theta \cos
  \phi \cos \psi - \sin \phi \sin \psi) \\ \upsilon ( \cos \theta \sin
  \phi \cos \psi + \cos \phi \sin \psi ) \\ -\upsilon \sin \theta \cos
  \psi \end{bmatrix},
\end{equation}
and $v$ is the rms velocity. Three angles, $\theta$, $\phi$ and
  $\psi$ correspond to the random selection of the string's
  orientation, whereas $\sigma$ and $\tau$ parameterise the string
  world sheet of straight segments.

\begin{figure*}[t]
\begin{center}
\includegraphics[scale=0.78]{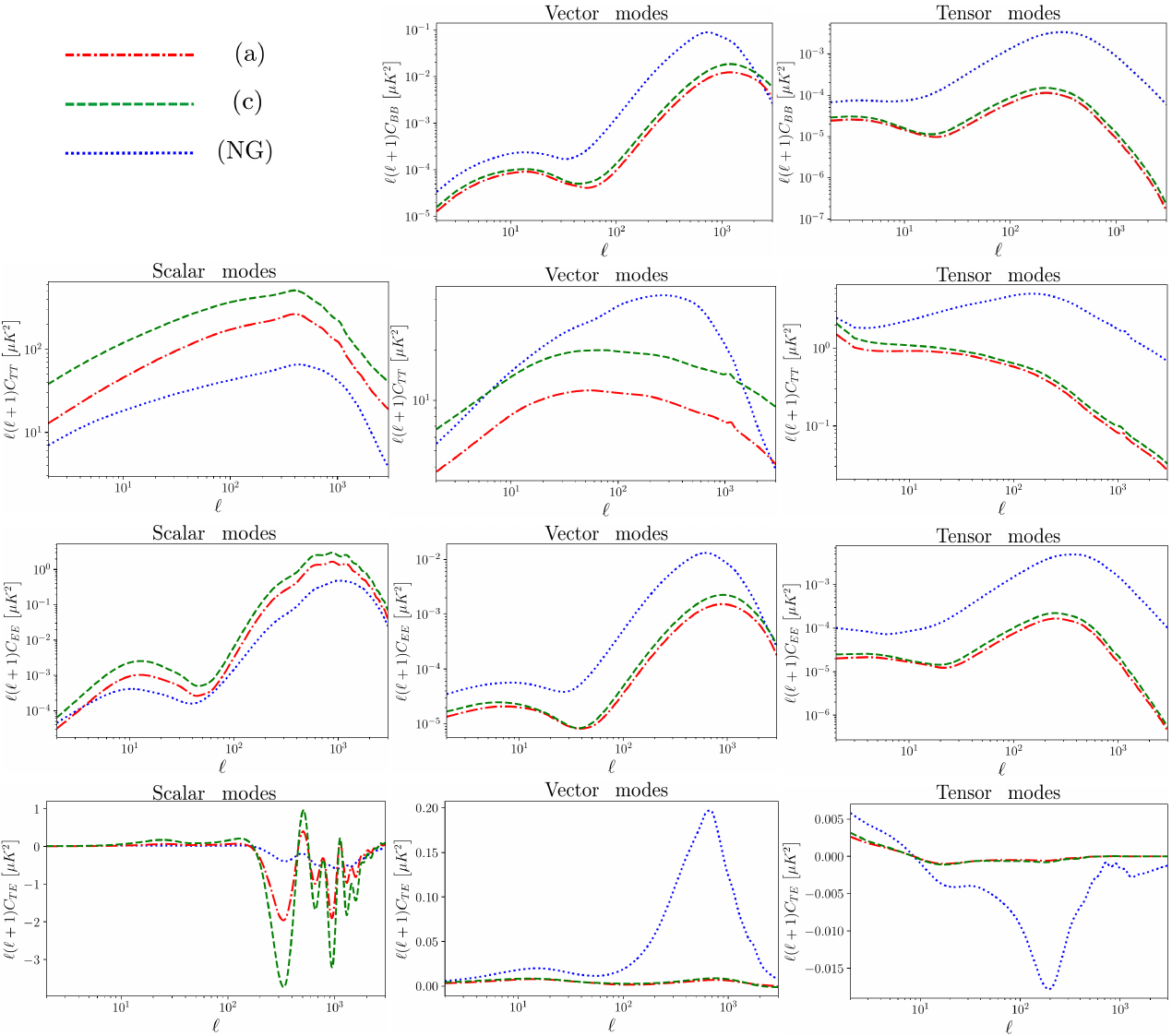}

\caption{\label{Fig:CMB_AC} CMB anisotropies arising from standard (NG) and current-carrying ``frozen'' cosmic string networks are investigated. The parameters align with the models (a) and (c) outlined in Sec. \ref{Other scenarios} and Ref.~\cite{Martins:2021cid}.}

\end{center}
\end{figure*}

We employ the macroscopic variables from the CVOS model, rather
  than the VOS model, to describe straight string segments in
  Eq.~\eqref{StrSeg}. As a result, the stress-energy tensor for the
  current-carrying string network can be expressed as the sum of the
  stress-energy tensors of all individual straight segments. The
  stress-energy tensor for each straight segment of a current-carrying
  string can be written using the macroscopic CVOS variables in the
  following way \cite{Rybak:2017yfu}
\begin{widetext}
\begin{equation}
    \label{SEtensor}
    T^{\mu \nu} = \frac{\mu_0}{\sqrt{-g}} \int \dd^2 \sigma
    \sqrt{-\gamma} \delta^{(4)} (x^{\lambda} - X^{\lambda}) \left[
      \Tilde{U} \Tilde{u}^{\mu} \Tilde{u}^{\nu} - \Tilde{T}
      \Tilde{v}^{\mu} \Tilde{v}^{\nu} - \Phi \left( \Tilde{u}^{\mu}
      \Tilde{v}^{\nu} + \Tilde{v}^{\mu} \Tilde{u}^{\nu} \right)
      \right],
\end{equation}
where we have defined $\Tilde{U} \equiv 1 + Y $, $\Tilde{T} \equiv 1 -
Y$, $\Phi \equiv \sqrt{ Y^2 - \frac14 K^2 } $, $\Tilde{u}^{\mu} \equiv
\dot{X}^{\mu}/ \sqrt{\dot{X}^2}$, and $\Tilde{v}^{\mu} \equiv
X^{\prime \, \mu} / \sqrt{-X^{\prime \, 2}}$. The Fourier-transformed
stress-energy tensor is given by
\begin{equation}
    \label{SEtensorF}
    \Theta^{\mu \nu} = \mu_0
    \int^{\xi_\textsc{c}/2}_{-\xi_\textsc{c}/2} \left[ \Tilde{U}
      \frac{\dot{X}^{\mu} \dot{X}^{\nu}}{\sqrt{1-v^2}} - \Tilde{T}
      \sqrt{1-v^2} X^{\prime \, \mu} X^{\prime \, \nu} - \Phi \left(
      \dot{X}^{\mu} X^{\prime \, \nu} + X^{\prime \, \mu}
      \dot{X}^{\nu} \right) \right] 
\text{e}^{i \bm{k}\cdot \bm{X}} \dd\sigma,
\end{equation}
where, without loss of generality, we use the wave vector oriented along the third component: $\bm{k} = k \bm{e}_3$.
Carrying out the integration in Eq.~\eqref{SEtensorF} one can obtain
expressions for the scalar $\Theta^\textsc{s} \equiv \frac12 \left( 2 \Theta_{33} - \Theta_{11} - \Theta_{22}\right)$, vector $\Theta^\textsc{v} \equiv \Theta_{13} $ and tensor $\Theta^\textsc{t} \equiv \Theta_{12}$ parts (see Sec.~II.C in Ref.~\cite{PhysRevD.59.023508} for details) in the following
form~\cite{Rybak:2017yfu}
\begin{subequations}
\label{Stress-Energy}
\begin{align}
   \label{Stress-EnergyChSc}
     \frac{\Theta^\textsc{s}}{\Theta_{00}} & = \frac12 \left[ v^2
       \left( 3 \dot{X}_3 \dot{X}_3 - 1 \right) - 6 v
       \frac{\Phi}{\Tilde{U}} X_3^{\prime} \dot{X}_3 -\left( 1-v^2
       \right) \frac{\Tilde{T}}{\Tilde{U}} \left( 3 X_3^{\prime}
       X_3^{\prime} -1 \right) \right],\\
   \label{Stress-EnergyChVect}    
      \frac{\Theta^\textsc{v}}{\Theta_{00}} & = v^2 \dot{X}_1
      \dot{X}_3 - \frac{\Tilde{T}}{\Tilde{U}} (1-v^2) X_1^{\prime}
      X_3^{\prime} - v \frac{\Phi}{\Tilde{U}} \left( X_1^{\prime}
      \dot{X}_3 + \dot{X}_1 X_3^{\prime} \right),\\
       \label{Stress-EnergyChTens}    
      \frac{\Theta^\textsc{t}}{\Theta_{00}} & = v^2 \dot{X}_1
      \dot{X}_2 - \frac{\Tilde{T}}{\Tilde{U}} (1-v^2) X_1^{\prime}
      X_2^{\prime} - v \frac{\Phi}{\Tilde{U}} \left( X_1^{\prime}
      \dot{X}_2 + \dot{X}_1 X_2^{\prime} \right),
\end{align}
\end{subequations}
where 
\begin{equation}
   \label{Stress-Energy00}
    \Theta_{00} = \frac{\mu_0 \Tilde{U}}{\sqrt{1-v^2}}
    \frac{\sin(\frac12 k X^{\prime}_3 \xi_\textsc{c})}{\frac12 k
      X^{\prime}_3} \cos(\bm{k} \cdot \bm{x}_0+k \dot{X}_3 v \tau).
\end{equation}
\end{widetext}
By expressing the contributions of individual string segments in terms of currents \ref{Stress-Energy}, one can perform summations across all segments to derive the complete stress-energy tensor for the cosmic string network under consideration. The string segments undergoing decay simultaneously are combined into what is referred to as ``consolidated string segments'' \cite{PogosianVachaspati}, as outlined in detail in Sec. III.A. of Ref.~\cite{PhysRevD.59.023508}.

Incorporating all the modifications into the \small{CMBACT} code, we can now make predictions for the CMB anisotropies arising from current-carrying cosmic strings.
One should notice that there is no assumption about scaling behavior
of the network. Therefore, we can apply this algorithm to a ``frozen''
network that might occur as a solution $v \propto \frac{\xi_\textsc{c}}{\tau} \propto 
\sqrt{1-Y} \propto \tau^{-s}$ as well as to the single-parameter scaling model to
which we now turn. References \cite{Martins:2021cid, Rybak:2023jjn} discuss
these solutions in detail.

\begin{figure*}[t]
\begin{center}
\includegraphics[scale=0.78]{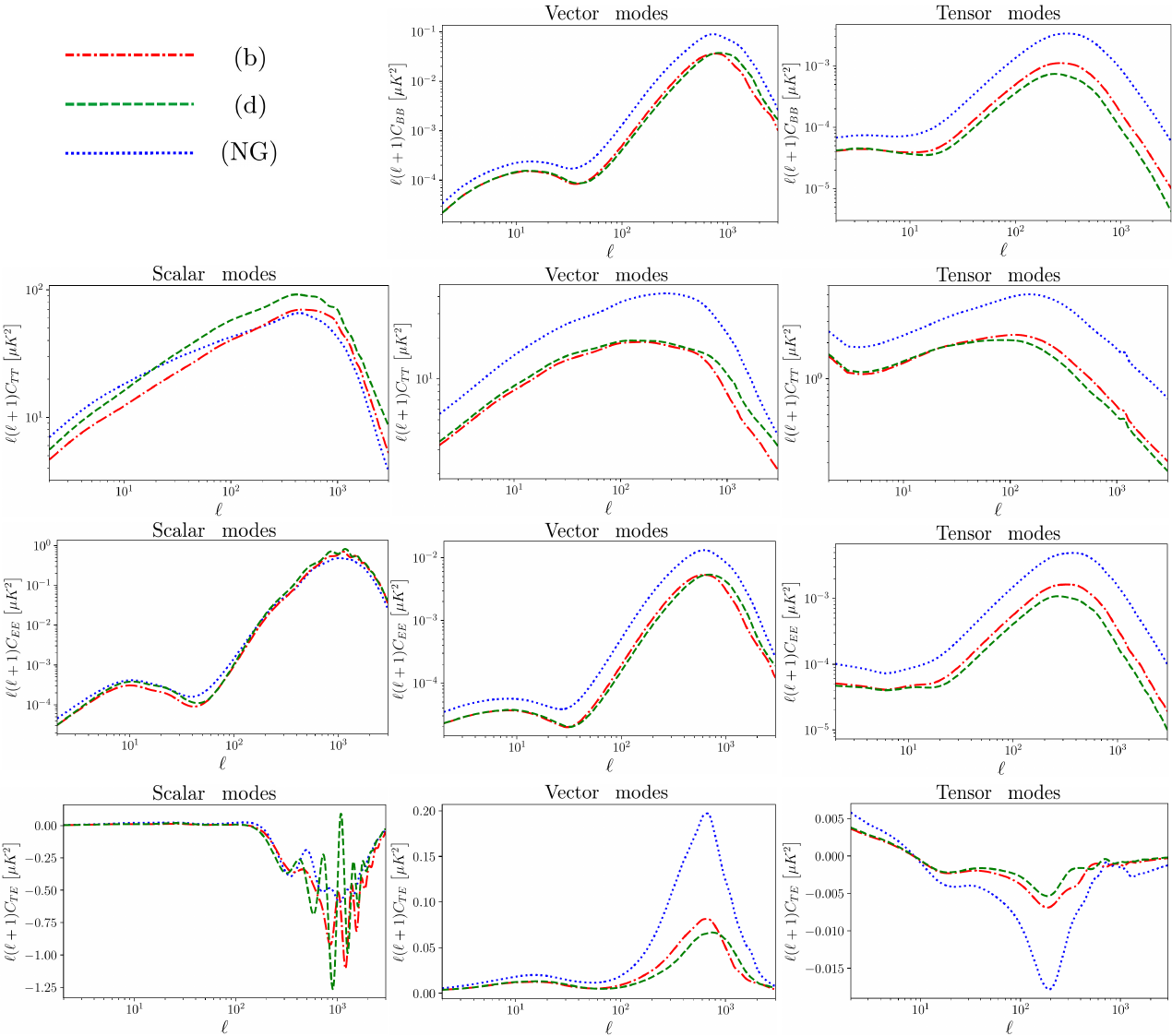}

\caption{\label{Fig:CMB_BD} CMB anisotropies originating from standard (NG) and current-carrying cosmic strings with the linear scaling regime are examined. The parameters are in accordance with the models (b) and (d) detailed in Sec. \ref{Other scenarios} and Ref.~\cite{Martins:2021cid}.}

\end{center}
\end{figure*}

For a detailed comparison of the CMB anisotropies originating from NG and current-carrying cosmic strings, we present figures that depict the separations for scalar, vector, and tensor modes for $BB$, $TT$, $EE$, and $TE$ components. Specifically, our analyses includes various values of $Y_{\rm sc}$, a parameter governing the charge value during the radiation-dominated period, as detailed in Sec. \ref{Ysc} and illustrated in Fig.~\ref{Fig:CMB_Y}.

Additionally, we extend our calculations to explore alternative scenarios. In Fig.~\ref{Fig:CMB_BD}, we show the results of CMB anisotropies generated by cosmic string networks with currents, exhibiting linear scaling regimes denoted as (b) and (d) models, as described in Sec. \ref{Other scenarios}. Furthermore, Fig.~\ref{Fig:CMB_AC} presents the CMB anisotropies stemming from cosmic string networks undergoing the ``frozen'' regime during the radiation-dominated epoch, characterized as (a) and (c) models, and described in Sec. \ref{Other scenarios}. 

Analysing Fig.~\ref{Fig:CMB_Y} one sees that, starting from the
NG limit ($Y_\textsc{sc}=0$) scalar modes have a slight
increase until $Y_\textsc{sc} \approx 0.3$, then decrease until
$Y_\textsc{sc} \approx 0.7$ and finally drastically increase for
$Y_\textsc{sc} > 0.7$. The vector and tensor modes in general have
monotonic decreases with increasing values of $Y_\textsc{sc}$. For
large multipole moments (small angular scales) $\ell \gtrsim 200$ for
vector modes we also see that after some threshold, the relevant
contribution increases.

This enhancement of scalar modes and decrease of vector and tensor
modes can be explained from the point of view of the string network
evolution. Specifically, the reason stems from the fact that with an
increase of the charge on cosmic strings, the rms velocity
decreases. This decrease of the velocity leads to the decrease of the
vector and tensor modes. At the same time, with smaller velocity the
string network loses comparatively less energy and therefore becomes
more dense, and a denser string network results in enhanced scalar
modes, which we also see for the (b) and (d) models in Fig.~\ref{Fig:CMB_BD}. 

The peak enhancement of scalar modes for current-carrying strings occurs when the network is in a frozen state ($Y_\textsc{sc}=1$). In both the (a) and (c) cases, the contribution from scalar modes significantly surpasses that of standard NG strings, as illustrated in the left column of Fig.~\ref{Fig:CMB_AC}. Simultaneously, the vector and tensor modes experience substantial suppression, as evident in the second and third columns of panels in Fig.~\ref{Fig:CMB_AC}. Notably, for small angular scales ($\ell>1000$), the vector modes of a frozen network dominate over the NG case. These distinctive features have the potential to serve as discriminators between ordinary cosmic strings and those involving current-carrying strings.

\bibliography{references}

\end{document}